\documentclass[pdftex,a4paper,12pt]{article}
\usepackage{amssymb,amsmath,amsfonts,ntheorem,nomencl,mathrsfs}

\usepackage[arrow, matrix, curve]{xy}
\usepackage[latin1]{inputenc}

\newcommand{\IC}{\mathbb{C}}
\newcommand{\IR}{\mathbb{R}}

\newcommand{\ILL}{\mathscr{L}}

\newcommand{\IHH}{\mathscr{H}}

\newcommand{\IFF}{\mathscr{F}}
\newcommand{\IAA}{\mathscr{A}}

\newcommand{\IN}{\mathbb{N}}

\newcommand{\pa}{\slash\slash}

\newcommand{\Id}{{\rm d}}

\newcommand{\f}{\frac}
\newcommand{\nn}{\nonumber}

\newtheorem{theorem}{THEOREM}[section]

\newtheorem{Lemma}[theorem]{Lemma}

\newtheorem{Corollary}[theorem]{Corollary}

\newtheorem{Remark}[theorem]{Remark}

\newtheorem{Theorem}[theorem]{Theorem}

\newtheorem{Proposition}[theorem]{Proposition}

\newtheorem{Definition}[theorem]{Definition}

\begin{document}
\begin{titlepage}
\title{On generalized Schrödinger semigroups }
 \author{Batu Güneysu\footnote{E-Mail: gueneysu@math.hu-berlin.de}\\
 Institut für Mathematik\\
 Humboldt-Universität zu Berlin\\}
\end{titlepage}

\maketitle 

\begin{abstract} We prove a Feynman-Kac formula for Schrödinger type operators on vector bundles over arbitrary Riemannian manifolds, where the potentials are allowed to have strong singularities, like those that typically appear in atomic quantum mechanical problems. This path integral formula is then used to prove several $\mathsf{L}^p$-type results, like bounds on the ground state energy and $\mathsf{L}^2 \leadsto\mathsf{L}^p$ smoothing properties of the corresponding Schrödinger semigroups. As another main result, we will prove that with a little control on the Riemannian structure, the latter semigroups are also $\mathsf{L}^2\leadsto \{\text{bounded continuous}\}$ smoothing for Kato decomposable potentials.
\end{abstract}

\section{Setting and some notation}\label{mani}

Let $M$ be a smooth Riemannian manifold (connected and without boundary), equipped with the Riemannian volume measure $\mathrm{vol}(\bullet)$. We set $m:=\dim M$ and denote the minimal positive heat kernel of $M$ with $p_t(x,y)$ and the scalar Laplace-Beltrami operator with $-\Delta=\Id^*\Id$. For $x,y\in M$, the number $\Id(x,y)$ will stand for the geodesic distance of $x,y\in M$ and $\mathrm{K}_r(x)$ for the open geodesic ball with radius $r$ around $x$.\\
Let $E\to M$ be a smooth (finite dimensional) complex vector bundle with a fixed Hermitian structure $(\bullet,\bullet)_x$ and a fixed Hermitian covariant derivative $\nabla$. The symbol $\left\|\bullet\right\|_x$ stands for the corresponding norm and operator norm of each fiber $E_x$. We will also use the notation 
\[
\left|\Psi\right|(x):=\left\|\Psi(x)\right\|_x\>\>\text{ for any section $\Psi$ in $E$ or in $\mathrm{End}(E)$.}
\]
The scalar product in $\Gamma_{\mathsf{L}^2}(M,E)$ will be written
\begin{align}
\left\langle f_1,f_2 \right\rangle =\int_M (f_1(x),f_2(x))_x \mathrm{vol}(\Id x)\label{xx3}
\end{align}                                                                                                             
and $\left\|\bullet\right\|$ stands for the norm and the operator norm corresponding to (\ref{xx3}). \vspace{1mm}

For our probabilistic considerations, we will assume that the underlying filtered probability space $(\Omega,\IFF,\IFF_*,\mathbb{P})$ satisfies the usual hypothesis and that it carries a Brownian motion $W$ in the Euclidean $\IR^l$, where $l\geq m$ is large enough. We will also assume $\IFF_*=\IFF_*(W)$. One can use this setting to construct a Brownian motion 
\[
B(x):[0,\zeta(x))\times\Omega \longrightarrow  M, 
\]
starting from $x$ with lifetime $\zeta(x)$, as the maximally defined solution of a Stratonovic\footnote{We will write $\underline{\Id}$ for Stratonovic differentials, whereas It\^{o} differentials will be written as $\Id$.} differential equation of the form
\begin{align}
\Id B(x) =\sum^l_{j=1} A_j(B(x)) \underline{\Id} W^j,\>\> B_0(x)=x,\label{dk2j}
\end{align}
where $A_1,\dots,A_l\in\Gamma_{\mathsf{C}^{\infty}}(M,\mathrm{T}M)$ are such that $\sum^l_{j=1} A^2_j=\Delta$. If $\pi:\mathrm{P}(E)\to M $ denotes the $\mathrm{U}(d)$ - principal bundle of unitary frames in $E$, then the stochastic $\nabla$-horizontal lift 
\[
U(u): [0,\zeta(x))\times\Omega \longrightarrow \mathrm{P}(E) 
\]
of $B(x)$ from a $\IFF_0$- random variable $u:\Omega\to\mathrm{P}(E)$ with $\pi(u)=x$ $\mathbb{P}$-a.s. is given as the maximally defined solution of
\[
\Id U(u) =\sum^l_{j=1} A^*_j(U(u)) \underline{\Id} W^j,\>\> U_0(u)=u,
\]
where $A^*_j\in \Gamma_{\mathsf{C}^{\infty}}(\mathrm{P}(E),\mathrm{T} \mathrm{P}(E))$ is the $\nabla$-lift of $A_j$. The fact that $U(u)$ indeed lives until $\zeta(x)$ follows from theorem 13C, p.175, in \cite{El}. The corresponding stochastic parallel transport will be written as an isometry 
\[
\pa^x_t:= U_t(u)u^{-1}: E_x \longrightarrow  E_{B_t(x)}\>\>\text{ $\mathbb{P}$-a.s. in $\{t<\zeta(x)\}$} 
\]
for any $t\geq 0$, with some $U(u)$ as above. As the notation indicates, the process $\pa^x$ does not depend on the particular choice of $u$ (see for example \cite{babba}, proposition 2.17). The reader may find the details of these constructions for example in \cite{Ha} \cite{babba} \cite{Gue} and the references therein. 

\section{Main results}

We will usually work under a global Kato assumption on some negative part of the potentials under consideration:

\begin{Definition}\label{sn2} A measurable function $v:M\to\IC$ is said to be in the Kato class $\mathcal{K}(M)$, if
\begin{align}
\lim_{t\searrow 0}\sup_{x\in M} \int^t_0 \int_M p_s(x,y) \left|v(y)\right| \mathrm{vol}(\Id y) \Id s= 0,
\end{align}
and $v$ is said to be in the local Kato class $\mathcal{K}_{\mathrm{loc}}(M)$, if $1_K v\in \mathcal{K}(M)$ for all compact $K\subset M$. 
\end{Definition}

In general, $\mathcal{K}(M)$ and also $\mathcal{K}_{\mathrm{loc}}(M)$ can depend on the Riemannian structure of $M$. Furthermore, using general properties of $p_t(x,y)$ one easily gets (\cite{batu}, proposition 2.7) the generally valid inclusions
\begin{align}
&\mathsf{L}^{\infty}(M)\subset\mathcal{K}(M),\>\>\mathcal{K}_{\mathrm{loc}}(M)\subset\mathsf{L}^1_{\mathrm{loc}}(M),
\end{align}
and with some control on the Riemannian structure of $M$, one can produce a large class of (local) Kato potentials:

\begin{Theorem}\label{Aa} Let $M$ be geodesically complete with Ricci curvature bounded from below and assume that there is a $C>0$ and a $R>0$ such that for all $0<r<R$ and all $x\in M$ one has $\mathrm{vol}(\mathrm{K}_r(x))\geq C r^{m}$. Then for any $p$ such that $p\geq 1$ if $m=1$, and $p > m/2$ if $m\geq 2$, one has 
\begin{align}
\mathsf{L}^p(M)+\mathsf{L}^{\infty}(M)\subset \mathcal{K}(M).\label{gjus}
\end{align}
In particular, for such $p$ one has $\mathsf{L}^p_{\mathrm{loc} }(M)\subset\mathcal{K}_{\mathrm{loc}}(M)$.
\end{Theorem}

{\it Proof.} See corollary 2.11 in \cite{batu} for first inclusion. The second inclusion is a trivial consequence of the first one. \vspace{0.5mm}

\hfill$\blacksquare$ 

\begin{Remark}\label{ggf}{\rm  If $M$ is geodesically complete with Ricci curvature bounded from below and a positive injectivity radius, then $M$ satisfies the assumptions of theorem \ref{Aa}. This is included in \cite{kt}, p.110.}
\end{Remark}

We will frequently make use of the following two compatibility results, that are valid without additional assumptions on the Riemannian structure of $M$.  

\begin{Lemma}\label{Katto0} {\rm a)} Let $v\in\mathsf{L}^1_{\mathrm{loc}}(M)$. Then for a.e. $x\in M$ one has 
\begin{align}
\mathbb{P}\left\{v(B_{\bullet}(x))\in\mathsf{L}^1_{\mathrm{loc}}[0,\zeta(x))\right\}=1. \label{qtm0}
\end{align}
{\rm b)} Let $v\in\mathcal{K}_{\mathrm{loc}}(M)$. Then for any $x\in M$ one has (\ref{qtm0}).
\end{Lemma}

{{\it Proof.} We will assume that $M$ is noncompact for the proof (the proof below can be easily adjusted to cover the compact case). Let $(K_n)$ be a relatively compact exhaustion of $M$ with domains $K_n\subset M$, and for any $x$ let $\zeta^{(1)}_n(x)$ be the first exit time of $B(x)$ from $K_n$. Since $B(x)$ has continuous paths, the sequence $\zeta^{(1)}_n(x)$ announces\footnote{that is, $\zeta^{(1)}_n(x)\nearrow\zeta(x)$ as $n\to\infty$ and $\zeta^{(1)}_n(x)<\zeta(x)$ for all $n$, $\mathbb{P}$-a.s. } $\zeta(x)$, so $\zeta^{(2)}_n(x):=\min(\zeta^{(1)}_n(x), n)$ also announces $\zeta(x)$. As a consequence, for any measurable $h:M\to \IC$ and any $j=1,2$ we have   
\begin{align}
&\mathbb{P}\left\{h(B_{\bullet}(x))\in\mathsf{L}^1_{\mathrm{loc}}[0,\zeta(x))\right\}=\mathbb{P} \bigcap_{n\in\IN}\left\{\int^{\zeta^{(j)}_n(x)}_0\left|h(B_s(x))\right|\Id s<\infty\right\}. \nn
\end{align}
We will also use the facts
\begin{align}
&\mathbb{E}\left[1_{\left\{s<\zeta(x)\right\}}\left|h(B_s(x))\right|\right]=\int_M p_s(x,y) \left|h(y)\right| \mathrm{vol}(\Id y),\>\>\int_M p_s(x,y)\mathrm{vol}(\Id y)\leq 1\nn
\end{align}
(valid for all $s>0$, $x\in M$) in the following.\\
a) Let us first assume that $v\in\mathsf{L}^1(M)$. Then, using Fubini, for any $n$ we have
\begin{align} 
&\int_M \mathbb{E}\left[   \int^{\zeta^{(2)}_n(x)}_0 \left|v(B_s(x)) \right|\Id s    \right] \mathrm{vol}(\Id x) \nn\\
&\leq \int_M \mathbb{E}\left[   \int^{\min(\zeta(x), n)}_0 \left|v(B_s(x)) \right|\Id s   \right] \mathrm{vol}(\Id x) \nn\\
& = \int_M \mathbb{E}\left[  \int^{n}_0 1_{ \{s<\zeta(x)\}  } \left|v(B_s(x)) \right| \Id s \right] \mathrm{vol}(\Id x)\nn\\
&= \int^n_0 \int_M  \int_M p_s(x,y)  \mathrm{vol}(\Id x) \left|v(y)\right|  \mathrm{vol}(\Id y)\Id s <\infty,\label{gtz}
\end{align}
which implies (\ref{qtm0}) in this situation. If one only has $v\in\mathsf{L}^1_{\mathrm{loc}}(M)$, then, since now $1_{K_n}v\in \mathsf{L}^1(M)$, for a.e. $x$ and all $n$ we have
\begin{align}
&\mathbb{P} \left\{\int^{\zeta^{(1)}_n(x)}_0\left|v(B_s(x))\right|\Id s=\infty\right\}\nn\\
=\>\>\>&\mathbb{P}
\left\{\int^{\zeta^{(1)}_n(x)}_0\left|\Big(1_{K_n}(B_s(x))+1_{M\setminus
K_n}(B_s(x))\Big)v(B_s(x))\right|\Id s=\infty\right\}\nn\\
\leq \>\>\> &\mathbb{P}\left\{\int^{\zeta^{(1)}_n(x)}_0\left|(1_{K_n}v)(B_s(x))\right|\Id
s=\infty\right\}=0,\label{hbv}
\end{align}
which again implies (\ref{qtm0}).\\
b) Let $x\in M$, $v\in\mathcal{K}(M)$, $n\in\IN$. We have 
\begin{align}
&\mathbb{E}\left[  \int^{\zeta^{(2)}_n(x)}_0 \left|v(B_s(x)) \right|\Id s  \right] \leq  \mathbb{E}\left[  \int^{\min(\zeta(x), n)}_0 \left|v(B_s(x)) \right|\Id s   \right] \nn\\
& =  \mathbb{E}\left[  \int^{n}_0 \left|v(B_s(x)) \right|1_{ \{s<\zeta(x)\}  }  \Id s   \right] = \int^n_0   \int_M p_s(x,y)  \left|v(y)\right|  \mathrm{vol}(\Id y) <\infty,
\end{align}
where the latter finiteness is trivial for small $n$ in view of the Kato property, and can then be extended to arbitrary $n$ using the Markoff Property of $B(x)$. This implies (\ref{qtm0}) for the global Kato case, and now one can use the same localization procedure as above to deduce (\ref{qtm0}) for arbitrary $v\in\mathcal{K}_{\mathrm{loc}}(M)$.\vspace{0.5mm}

\hfill$\blacksquare$\vspace{1.1mm}

\begin{Proposition}\label{xdd} For any $v\in\mathcal{K}(M)$ there is a $C(v)>0$ such that for all $t\geq 0$,
\[
\sup_{x\in M} \mathbb{E}\left[\mathrm{e}^{\int^t_0 \left|v(B_s(x))\right|\Id s}1_{\{t<\zeta(x)\}}\right]\leq  2 \mathrm{e}^{tC(v)}. 
\]
\end{Proposition}

{\it Proof.} Let $\hat{M}=M\cup\{\infty_M\}$ be the Alexandroff compactification of $M$. We can extend any measurable $w:M\to\IC$ to a function $\hat{w}:\hat{M}\to\IC$ by setting $\hat{w}(\infty_M)=0$, and we can also extend $B(x)$ to a process $\hat{B}(x):[0,\infty)\times \Omega\to \hat{M}$ by setting $\hat{B}_t(x)(\omega):=\infty_M$, if $t\geq \zeta(x)(\omega)$. Then one has 
\[
\mathbb{E}\left[\mathrm{e}^{\int^t_0 \left|v(B_s(x))\right|\Id s}1_{\{t<\zeta(x)\}}\right]\leq  \mathbb{E}\left[\mathrm{e}^{\int^t_0 \left|\hat {v}(\hat{B}_s(x))\right|\Id s}\right].
\]
Let
\[
C(v,s):= \sup_{x\in M} \mathbb{E}\left[\int^s_0\left|\hat{v}(\hat{B}_r(x))\right|\Id r\right]= \sup_{x\in M} \mathbb{E}\left[\int^s_0 |v(B_r(x))|1_{\{r<\zeta(x)\}}\Id r\right] 
\]
and choose a $t_0(v)>0$ with $C(v,t_0(v))<1/2$. Then using Khas'minskii's lemma and the Markoff property of $\hat{B}(x)$ one gets (see for example p.9 in \cite{Sz} for the arguments) the first inequality in
\begin{align}
\mathbb{E}\left[\mathrm{e}^{\int^t_0 \left|\hat{v}(\hat{B}_s(x))\right|\Id s}\right]&\leq \f{1}{1-C(v,t_0(v))} \mathrm{e}^{ \f{t}{t_0(v)} \mathrm{log}\left(\f{1}{1-C(v,t_0(v))}\right) }\nn\\
& < 2 \ \mathrm{e}^{ \f{t}{t_0(v)} \mathrm{log}\left(  \f{1}{1-C(v,t_0(v))}  \right) }.\nn
\end{align}
This proves the claim.

 \vspace{0.5mm}

\hfill$\blacksquare$\vspace{1.1mm}

We refer the reader to \cite{batu} and the references therein for more facts about Kato potentials on Riemannian manifolds.\\
We return to the operator setting: The operator $\nabla^*\nabla/2$ with domain of definition $\Gamma_{\mathsf{C}^{\infty}_0}(M,E)$ is a nonnegative symmetric operator in $\Gamma_{\mathsf{L}^2}(M,E)$ and the corresponding Friedrichs realization will be denoted with $H(0)\geq 0$. Since there won't be any danger of confusion, we will denote the Friedrichs realization of $-\Delta/2$ in $\mathsf{L}^2(M)$ again with the same symbol $-\Delta/2\geq 0$. The corresponsing quadratic forms in $\Gamma_{\mathsf{L}^2}(M,E)$ and in $\mathsf{L}^2(M)$, respectively, will be written as $q_{H(0)}$ and $q_{-\Delta/2}$.\vspace{1.2mm}

{\it Throughout, let 
\[
V:M\longrightarrow \mathrm{End}(E) 
\]
be a potential in the following}\footnote{By ``potential'' we mean a measurable section $V$ in $\mathrm{End}(E)$ such that $V(x)$ is Hermitian for almost every (a.e.) $x\in M$. }. \vspace{1.2mm}

Then one can define a quadratic form $q_{V}$ in $\Gamma_{\mathsf{L}^2}(M,E)$ as follows:
\begin{align}
&\mathsf{D}(q_{V})=\left.\Big\{f\right|f\in \Gamma_{\mathsf{L}^2}(M,E),\> \left( Vf,f\right)\in \mathsf{L}^1(M)\Big\},\nn\\
&q_{V}(f)= \int_M \left( V(x)f(x),f(x)\right)_x\mathrm{vol}(\Id x).
\end{align}
It will be convinient to introduce the notation 
\[
\underline{V}:M\longrightarrow \IR,\>\> \underline{V}(x):=\min\sigma(V(x)).
\]

{\it We also fix a scalar potential 
\[
v:M\longrightarrow \IR 
\]
in the following.}

\vspace{1mm}

The following theorem follows directly from theorem 2.13 in \cite{batu}:

\begin{Theorem}\label{hh2} Let $V$ be such that there is a decomposition $V=V^{(1)}-V^{(2)}$ into potentials $V^{(1)},V^{(2)}\geq 0$ with
\[
\left|V^{(1)}\right|\in\mathsf{L}^1_{\mathrm{loc}}(M)\>\text{ and }\>\left|V^{(2)}\right|\in\mathcal{K}(M).
\]
Then one has
\[
\mathsf{D}(q_{H(0)}+q_{V}):=\mathsf{D}(q_{H(0)})\cap \mathsf{D}(q_{V})=\mathsf{D}(q_{H(0)})\cap \mathsf{D}(q_{V^{(1)}})
\]
and $q_{H(0)}+q_{V}$ is a densely defined, closed and semibounded from below quadratic form in $\Gamma_{\mathsf{L}^2}(M,E)$. 
\end{Theorem}

\begin{Remark}{\rm Note that the above decomposition of $V$ into nonnegative potentials need not be the canonic one $V=V^+-V^-$ which comes from the fiberwise spectral calculus of $E$.}
\end{Remark}

It follows from theorem \ref{hh2} that the form sum $H(0) \dotplus V$ (= the operator corresponding to $q_{H(0)}+q_{V}$) is a well-defined self-adjoint semibounded from below operator in $\Gamma_{\mathsf{L}^2}(M,E)$ which will be denoted with $H(V)$. Generalizing the situation considered in \cite{Si0}, we will call 
\[
(\mathrm{e}^{-t H(V)})_{t\geq 0}\subset \ILL(\Gamma_{\mathrm {L}^2}(M,E)) 
\]
the {\it Schrödinger semigroup} corresponding to $H(V)$. 

\begin{Remark}\label{mav}{\rm 1. We use the following notation for scalar operators on functions: If $\beta\in\Omega^1_{\IR}(M)$ and if $v$ is such that there is a decomposition $v=v^1-v^{(2)}$ with $0\leq v^{(1)}\in\mathsf{L}^1_{\mathrm{loc}}(M)$ and $0\leq v^{(2)}\in\mathcal{K}(M)$, then the self-adjoint semibounded from below operator in $\mathsf{L}^2(M)$ corresponding to $(\Id+\mathrm{i}\beta)^*(\Id+\mathrm{i}\beta)/2+v$ in the sense of theorem \ref{hh2} (applied to $\nabla=\Id+\mathrm{i}\beta$) will be written as $H_{\beta}(v)$, with the convention $H_0(0)=-\Delta/2$. Operators of the form $H_{\beta}(v)$ describe the energy of charged nonrelativistic quantum mechanical particles with spin $0$, which live on $M$ under the influence of the potential $v$ and the magnetic field $\Id\beta$. \vspace{1.2mm}

2. The above smoothness assumption on the magnetic potential $\beta$ is satisfactory from the physics point of view, since, at least, this is a local assumption. The above class of potentials $v$, on the other hand, is certainly big enough to deal with most physically relevant situations. This claim is motivated by (\ref{gjus}), which implies that the Kato class is big enough to deal with Coulomb type singularities $-1/|x|_{\IR^3}$ in the Euclidean $\IR^3$, which appear naturally in the quantum mechanical hydrogen problem. Similar Hydrogen type problems can also be considered on (nonparabolic) Riemannian manifolds \cite{enciso}\cite{G6}.}
\end{Remark}

Let us now state theorem \ref{sa2}, a scalar Feynman-Kac formula for Schrödinger operators of the form $H_0(v)$. We have prefered to first treat the scalar case seperately for two reasons: Firstly, the proof of theorem \ref{sa2} serves as a model for the proof of the Feynman-Kac formula for generalized operators of the type $H(V)$ (theorem \ref{hb3}), and secondly it is interesting to see that one can even {\it use} theorem \ref{sa2} itself applied to $H_0(\underline{V})$ for a convergence argument in the proof theorem \ref{hb3}. The latter ``scalarization procedure'' reflects the fact that operators of the form $H(V)$ always dominate scalar operators of the form $H_0(\underline{V})$, a statement which can be made precise by means of a Kato type inequality \cite{Br}\cite{batu}. We will derive and use several aspects of this domination in this paper.

\begin{Theorem}\label{sa2} Let $v$ be such that there is a decomposition $v=v^{(1)}-v^{(2)}$ with $0\leq v^{(1)}\in\mathsf{L}^1_{\mathrm{loc}}(M)$ and $0\leq v^{(2)}\in\mathcal{K}(M)$. Then for a.e. $x\in M$ one has 
\begin{align}
\mathbb{P}\left\{v(B_{\bullet}(x))\in\mathsf{L}^1_{\mathrm{loc}}[0,\zeta(x))\right\}=1, \label{xx22}
\end{align}
and the following formula holds for any $f\in\mathsf{L}^2(M)$, $t\geq 0$ and a.e. $x\in M$,
\begin{align}
\mathrm{e}^{-t H_0(v)}f(x)=\mathbb{E}\left[\mathrm{e}^{-\int^t_0 v(B_s(x))\Id s}f(B_t(x))1_{\{t<\zeta(x)\}}\right].
\label{zz4}
\end{align}
\end{Theorem}

Note that we do not make any assumptions on the Riemannian structure of $M$. The proof of theorem \ref{sa2} will be given in section \ref{scall}. \vspace{1.1mm}

We will use the notation 
\[
\mathscr{E}_{H}:=\min \sigma (H) 
\]
for the ground state energy of a self-adjoint semi-bounded from below operator $H$. It follows from theorem \ref{sa2} that $(\mathrm{e}^{-t H_0(v)})_{t>0}$ is positivity improving. Using this fact, we get the following facts for $\mathscr{E}_{H_0(v)}$ directly from abstract results on self-adjoint semi-bounded from below operators on measure spaces: 

\begin{Corollary} Fix the assumptions of theorem \ref{sa2}.\vspace{1.2mm}

{\rm a)} If $\mathscr{E}_{H_0(v)}$ is an eigenvalue of $H_0(v)$, then $\mathscr{E}_{H_0(v)}$ is simple and the corresponding ground state eigenfunction can be chosen strictly positive. \vspace{1.2mm}

{\rm b)} Let $f_1,f_2\in\mathsf{L}^2(M)\setminus\{0\}$ with $f_1,f_2\geq 0$. Then the following formula holds, 
\begin{align}
\mathscr{E}_{H_0(v)}= -\lim_{t\to\infty} t^{-1} \log \mathbb{E}\left[\int_M 1_{\{t<\zeta(x)\}} f_1(x) \mathrm{e}^{-\int^t_0 v(B_s(x))\Id s}f_2(B_t(x))\mathrm{vol}(\Id x)\right].\label{cc2}
\end{align}
\end{Corollary}

{\it Proof.} a) This follows directly from the fact that $(\mathrm{e}^{-t H_0(v)})_{t>0}$ is positivity improving. See for example theorem XIII.44 in \cite{Re4}. \vspace{1.2mm}

b) Using again that $(\mathrm{e}^{-t H_0(v)})_{t>0}$ is positivity improving, one has (see for example theorem 2.2 in \cite{lenz}) 
\begin{align}
\mathscr{E}_{H_0(v)}=-\lim_{t\to\infty}  \f{\log\left\langle f_1,\mathrm{e}^{-t H_0(v)}f_2\right\rangle}{t}.      \label{c33}
\end{align}
Now (\ref{cc2}) follows from (\ref{c33}) by the Feynman-Kac formula and Fubini's theorem. \vspace{0.5mm}

\hfill$\blacksquare$\vspace{1.1mm}

We return to the general vector valued setting again. If $x\in M$ is appropriate, then the process
\[
\mathscr{V}^{x}:[0,\zeta(x))\times \Omega \longrightarrow\mathrm{End}(E)_x
\] 
will stand for the unique pathwise weak solution of 
\begin{align}
\Id  \mathscr{V}^{x}_{t} = -\mathscr{V}^{x}_{t}\Big( \pa^{x,-1}_t V(B_t(x)) \pa^{x}_t\Big)\Id t,\>\mathscr{V}^{x}_{0}=\mathbf{1}.\label{gg6}
\end{align}
Then $\mathscr{V}^{x}$ is pathwise invertible and 
\[
\mathscr{V}^{x,-1}:[0,\zeta(x))\times \Omega \longrightarrow\mathrm{End}(E)_x
\] 
is uniquely determined by
\begin{align}
\Id  \mathscr{V}^{x,-1}_t = \Big( \pa^{x,-1}_t V(B_t(x)) \pa^{x}_t\Big) \mathscr{V}^{x,-1}_t\Id t,\>\mathscr{V}^{x,-1}_{0}=\mathbf{1}. 
\end{align}

The following Feynman-Kac type formula for sections in $E$ will be the main tool of this paper. It is a generalization of theorem 1.3 in \cite{Gue} to not necessarily (geodesically or stochastically) complete $M$'s and to $V$'s that are not necessarily locally square integrable or bounded from below:

\begin{Theorem}\label{hb3} Let $V$ be such that there is a decomposition $V=V^{(1)}-V^{(2)}$ into potentials $V^{(1)},V^{(2)}\geq 0$ with 
\[
\left|V^{(1)}\right|\in\mathsf{L}^1_{\mathrm{loc}}(M)\>\text{ and }\>\left|V^{(2)}\right|\in\mathcal{K}(M).
\]
Then for a.e. $x\in M$, there is a unique process 
\[
\mathscr{V}^{x}:[0,\zeta(x))\times \Omega \longrightarrow\mathrm{End}(E)_x
\] 
which satisfies (\ref{gg6}) pathwise in the weak sense, and for any $f\in \Gamma_{\mathsf{L}^2}(M,E)$, $t\geq 0$, a.e. $x\in M$ one has
\begin{align}
\mathrm{e}^{-t H(V) }f(x)= \mathbb{E} \left[  \mathscr{V}^{x}_t \pa_t^{x,-1} f(B_t(x))1_{\{t<\zeta(x)\}}\right].\label{ii7}
\end{align}
\end{Theorem}

{\it Proof.} Firstly, we remark that since parallel transport is an isometric operation, the asserted existence of $\mathscr{V}^{x}$ will follow from the Banach fixed point theorem, if we can show that for a.e. $x\in M$ one has 
\begin{align}
\mathbb{P}\left\{\left\|V(B_{\bullet}(x))\right\|_{B_{\bullet}(x)}\in\mathsf{L}^1_{\mathrm{loc}}[0,\zeta(x))\right\}=1.\label{s88}
\end{align}
But this follows from the assumptions on $V$ and lemma \ref{Katto0}.\\
As in the proof of theorem \ref{sa2}, we divide the proof into two parts again:\vspace{1.2mm}

I) {\it (\ref{ii7}) holds under the additional assumption $V\geq C$.}\vspace{1.2mm}

Proof: We may assume $V\geq 0$. Using the the spectral calculus of the fibers of $E$ we define $V_n:=\min(n,V)$ for any $n\in\IN$. Then each $V_n$ is a potential with $|V_n|\in\mathsf{L}^{\infty}(M)$ and one has 
\begin{align}
0\leq V_n\leq V_{n+1}\leq V,\>V_n\to V \>\>\text{ a.e. in $M$ as $n\to\infty$.} \label{jh44}
\end{align}
Using monotone convergence of quadratic forms as in the first part of the proof of theorem \ref{sa2} shows that we may assume   
\begin{align}
\lim_{n\to\infty}\mathrm{e}^{-t H(V_n)}f(x)= \mathrm{e}^{-t H(V)}f(x)\>\> \text{  for a.e. $x$.} \label{d1}
\end{align}
With an obvious notation, proposition \ref{besch} implies
\begin{align}
\mathrm{e}^{-t H(V)}f(x)=\lim_{n\to\infty}\mathbb{E} \left[  \mathscr{V}^{x}_{n,t} \pa_t^{x,-1} f(B_t(x))1_{\{t<\zeta(x)\}}\right]\>\> \text{  for a.e. $x$.}\label{d2}
\end{align}

Let $x$ be such that (\ref{s88}) holds from now on. In view of (\ref{s88}) and (\ref{jh44}), proposition \ref{schlesi}  implies\footnote{Note that $|V_n|(\bullet)\leq |V|(\bullet)$, which follows from $\left|V_{n}\right|(\bullet)=\max \sigma(V_{n}(\bullet))$, $\left|V\right|(\bullet)=\max \sigma(V(\bullet))$ and (\ref{ap0}).}
\begin{align}
&\left\| \mathscr{V}^{x}_{n,t} -\mathscr{V}^{x}_{t} \right\|_x 1_{\{t<\zeta(x)\}}\nn\\
&\leq \mathrm{e}^{3\int^t_0\left\| V(B_s(x)) \right\|_{B_s(x)}\Id s}\int^t_0 \left\|V(B_s(x)) - V_n(B_s(x))  \right\|_{B_s(x)}\Id s 1_{\{t<\zeta(x)\}}\>\>\text{$\mathbb{P}$-a.s.},\label{d5}
\end{align}
so using (\ref{s88}) and (\ref{jh44}) again, we get from dominated convergence  that
\begin{align}
\lim_{n\to\infty}\left\| \mathscr{V}^{x}_{n,t} -\mathscr{V}^{x}_{t} \right\|_x1_{\{t<\zeta(x)\}}=0 \>\>\text{$\mathbb{P}$-a.s.}\label{f5}
\end{align}
Finally, we may use (\ref{f5}) and $\left\| \mathscr{V}^{x}_{n,t} \right\|_x1_{\{t<\zeta(x)\}}\leq 1$ $\mathbb{P}$-a.s. (the latter follows from $V_n\geq 0$ and proposition \ref{absc} c)), to deduce (\ref{ii7}) from (\ref{d2}) and dominated convergence. \vspace{1.2mm}

II) {\it (\ref{ii7}) holds in the general case.}\vspace{1.2mm}

Proof: Now we define $V_n:=\max(-n,V)$ for any $n$. Then each $V_n$ is a bounded from below locally integrable potential and one has
\begin{align}
V_{n}\geq V_{n+1}\geq V,\>V_n \to V\text{ a.e. in $M$ as $n\to\infty$,}\label{d67}
\end{align}
so that one can use convergence of monotonely decreasing quadratic forms as in the second part of the proof of theorem \ref{sa2} to see that we can assume (\ref{d1}). By I), we also have (\ref{d2}) now, and so it remains to prove that the limit may be put into the expectation value in (\ref{d2}) for a.e. $x$, which will be proved with a dominated convergence argument. To this end, note that we again have (\ref{d5}) and that (\ref{d67}) implies\footnote{ To see this inequality, just note $\left|V_{n}-V \right|(\bullet)=\max \sigma(V_{n}(\bullet)-V(\bullet))$ and use (\ref{ap0}).}
\begin{align}
0\leq |V_{n}-V|\leq |V_{1}-V|.
\end{align}
As a consequence, we may use theorem 12.2.6 in \cite{Jo} to deduce (\ref{f5}). Next, the inequality\footnote{ This follows directly from (\ref{ap0}).} $-\underline{V_n}\leq \> - \underline{V}$ and proposition \ref{absc} c) give 
\[
\left\| \mathscr{V}^{x}_{n,t} \right\|_x1_{\{t<\zeta(x)\}}\leq \mathrm{e}^{-\int^t_0\underline{V}(B_s(x))\Id s}1_{\{t<\zeta(x)\}}\>\>\text{ $\mathbb{P}$-a.s.,}\label{bhu}
\]
in particular,
\[
\left\| \mathscr{V}^{x}_{n,t}\pa_t^{x,-1} f(B_t(x)) \right\|_x1_{\{t<\zeta(x)\}}\leq \mathrm{e}^{-\int^t_0\underline{V}(B_s(x))\Id s}\left\|  \ f(B_t(x))\right\|_{B_t(x)}1_{\{t<\zeta(x)\}}
\]
$\mathbb{P}$-a.s. These arguments are valid for any $x$ such that (\ref{s88}) holds. Finally, we have 
\begin{align}
&\mathbb{E} \left[\mathrm{e}^{-\int^t_0\underline{V}(B_s(x))\Id s} \left\| f(B_t(x))\right\|_{B_t(x)}1_{\{t<\zeta(x)\}}\right]\nn\\
&=\mathrm{e}^{-t H_0(\underline{V})}|f|(x)<\infty\>\>\text{ for a.e. $x$},
\end{align}
since the scalar potential 
\[
\underline{V}(x)=\min\sigma(V(x))=\min\sigma(V^{(1)}(x))-\max\sigma(V^{(2)}(x))
\]
satisfies the assumptions of theorem \ref{sa2}, so that formula (\ref{ii7}) indeed follows from dominated convergence. \vspace{0.5mm}

\hfill$\blacksquare$\vspace{1.1mm}

Using the obvious extension of proposition 2.6 in \cite{Gue} to possibly incomplete $M$'s, one can immediately derive a very general Feynman-Kac-It\^{o} formula for magnetic Schrödinger operators on Riemannian manifolds from formula (\ref{ii7}):

\begin{Corollary}\label{itt} Let $v$ be such that there is a decomposition $v=v^{(1)}-v^{(2)}$ with $0\leq v^{(1)}\in\mathsf{L}^1_{\mathrm{loc}}(M)$ and $0\leq v^{(2)}\in\mathcal{K}(M)$, and let $\beta\in\Omega^1_{\IR}(M)$. Then the following formula holds for any $f\in\mathsf{L}^2(M)$, $t\geq 0$ and a.e. $x\in M$,
\begin{align}
\mathrm{e}^{-t H_{\beta}(v)}f(x)=\mathbb{E}\left[\mathrm{e}^{-\int^t_0 v(B_s(x))\Id s+\mathrm{i}\int^t_0\beta(\underline{\Id} B_s(x))}f(B_t(x))1_{\{t<\zeta(x)\}}\right],\label{dcdc}
\end{align}
where 
\[
\int\beta(\underline{\Id} B(x)):[0,\zeta(x))\times\Omega\longrightarrow \IR 
\]
stands for the Stratonovic integral of $\beta$ along $B(x)$.
\end{Corollary}

Formula (\ref{dcdc}) generalizes the Feynman-Kac-It\^{o} formula from corollary 1.5 in \cite{Gue} to possibly incomplete $M$'s and to more general $v$'s.\vspace{1.2mm}

Now we would like present some applications of (\ref{gg6}). We first come to some $\mathsf{L}^p$-type results. A key observation is the following semigroup domination. We refer the reader to \cite{Hess1} for an abstract formulation of semigroup domination and its applications. 

\begin{Theorem}\label{qp} Under the assumptions of theorem \ref{hb3}, let $v$ be such that there is decomposition $v=v^{(1)}-v^{(2)}$ with $0\leq v^{(1)}\in \mathsf{L}^1_{\mathrm{loc}}(M)$, $0\leq v^{(2)}\in\mathcal{K}(M)$ and assume furthermore that $V\geq v\mathbf{1}$. Then the following inequality holds for any $f\in\Gamma_{\mathsf{L}^2}(M,E)$, $t\geq 0$ and a.e. $x\in M$,
\begin{align}
\left\|\mathrm{e}^{-t H(V)}f(x)\right\|_x\leq \mathrm{e}^{-t H_0(v)}\left|f\right|(x).\label{api}
\end{align}
In particular, the following assertions hold:
\begin{itemize}
\item[\rm{i)}] For any $f\in\Gamma_{\mathsf{L}^2}(M,E)$, $t\geq 0$,
\begin{align}
\left\langle \mathrm{e}^{-t H(V)}f,f\right\rangle\leq \left\langle \mathrm{e}^{-t H_0(v)}\left|f\right|,\left|f\right|\right\rangle.\label{unn}
\end{align}
\item[\rm{ii)}] One has $\left|f\right|\in \mathsf{D}(q_{H_0(v)})$ with $q_{H(V)}(f)\geq q_{H_0(v)}(|f|)$ for any $f\in  \mathsf{D}(q_{H(V)})$.
\item[\rm{iii)}] One has $\mathscr{E}_{H(V)}\geq \mathscr{E}_{H_0(v)}$.
 \item[\rm{iv)}] For any $f\in\Gamma_{\mathsf{L}^2}(M,E)$, $k\in\IN$, $\lambda\in\IC$ with $\mathrm{Re}(\lambda )>\mathscr{E}_{H(V)}$, and a.e. $x\in M$,
\begin{align}
 \left\| (H(V)+\lambda)^{-k}f(x)\right\|_x\leq (H_0(v)+\lambda)^{-k}\left|f\right|(x).\label{reso}
\end{align}
\end{itemize}
\end{Theorem}

{\it Proof.} It follows from proposition \ref{absc} c) that for a.e.  $x\in M$ one has
\begin{align}
\left\|\mathscr{V}^{x}_t\right\|_x1_{\{t<\zeta(x)\}}\leq \mathrm{e}^{-\int^t_0 v(B_s(x))\Id s}1_{\{t<\zeta(x)\}}\>\>\text{  $\mathbb{P}$-a.s.,}\label{dhj}
\end{align}
which directly implies (\ref{api}) and (\ref{unn}) in view of the Feynman-Kac formulae. \\
For ii) and iii), we can assume that $H(V)$ and $H_0(v)$ are nonnegative (otherwise we can consider $H(V+C)$ and $H_0(v+C)$ with $C\in\IR$ small enough and use (\ref{ap2}) and (\ref{ap3})). Under this assumption, ii) is implied by i), (\ref{ap4}), and iii) follows from ii), (\ref{ap2}).\\
Finally, (\ref{reso}) follows from (\ref{api}) by taking the Laplace transforms
\[
(H(V)+\lambda)^{-k}f=\f{1}{(k-1)!}\int^{\infty}_0 t^{k-1} \mathrm{e}^{-t\lambda} \mathrm{e}^{-t H(V)} f \Id t
\]
and
\[
(H_0(v)+\lambda)^{-k}\left|f\right|=\f{1}{(k-1)!}\int^{\infty}_0 t^{k-1} \mathrm{e}^{-t\lambda} \mathrm{e}^{-t H_0(v)} \left|f\right| \Id t,
\]
and the proof is complete.\vspace{0.5mm}

\hfill$\blacksquare$\vspace{1.1mm}

\begin{Remark}{\rm 1. A canonical choice for $v$ in theorem \ref{qp} is given by $v:=\underline{V}$.\vspace{1mm}

2. If $V\geq 0$ is locally integrable, then theorem \ref{qp} iii) implies $\mathscr{E}_{H(V)}\geq \mathscr{E}_{-\Delta/2}$. In case $\mathscr{E}_{-\Delta/2}>0$, the latter inequality produces a nontrivial lower bound on the ground state energy of $H(V)$ which is  purely ``Riemann geometric`` in the sense that it does not depend on the interaction $V$ or any data corresponding to the underlying vector bundle $E$.}
\end{Remark}

Combining theorem \ref{qp} iii) with remark \ref{mav} leads to an important consequence for (nonrelativistic) quantum mechanics on Riemannian manifolds which is well-known for quantum mechanics in Euclidean space:

\begin{Corollary}\label{dsu} Let $v$ be such that there is a decomposition $v=v^{(1)}-v^{(2)}$ with $0\leq v^{(1)}\in \mathsf{L}^1_{\mathrm{loc}}(M)$, $0\leq v^{(2)}\in\mathcal{K}(M)$. Then the presence of a magnetic field with potential $\beta\in\Omega^1_{\IR}(M)$ leads to an increase of the ground state energy of charged nonrelativistic spin $0$  particles, which live on $M$ under the influence of $v$ and $\Id \beta$.
\end{Corollary}

{\it Proof.} Mathematically, the assertion just means that 
\begin{align}
\mathscr{E}_{H_{\beta}(v)}\geq \mathscr{E}_{H_0(v)},\label{nvzh}
\end{align}
and this inequality follows directly from theorem \ref{qp} iii). \\
However, we find it instructive to remark that it is almost trivial to deduce (\ref{nvzh}) directly from the Feynman-Kac-It\^{o} formula: For (\ref{dcdc}) and the triangle inequality implies
\begin{align}
\left\langle \mathrm{e}^{-t H_{\beta}(v)}f,f \right\rangle_{\mathsf{L}^2(M)}\leq \left\langle \mathrm{e}^{-t H_0(v)}\left|f\right|,\left|f\right| \right\rangle_{\mathsf{L}^2(M)}
\end{align}
for any $f\in\mathsf{L}^2(M)$, from which (\ref{nvzh}) follows directly from combining (\ref{ap0}) with  (\ref{ap3}).  \vspace{0.5mm}

\hfill$\blacksquare$\vspace{1.1mm}

We continue with our main results. For the next proposition we consider $\mathrm{e}^{-t H(V)}f$ and $\mathrm{e}^{-t H_0(v)}h$, where $f\in\Gamma_{\mathsf{L}^2}(M,E)$, $h\in\mathsf{L}^2(M)$, as an equivalence class of measurable sections in $E$ and, respectively, as an equivalence class of measurable functions on $M$. In this sense, both $\mathrm{e}^{-t H(V)}f$ and $\mathrm{e}^{-t H_0(v)}h$ are given by the corresponding Feynman-Kac formula. For any $p,q\in[1,\infty]$ let $\left\| \bullet\right\|_q$ denote the norm in $\Gamma_{\mathsf{L}^q}(M,E)$ and let $\left\|\bullet\right\|_{p,q}$ denote the norm corresponding to 
\[
\text{the Banach space $\ILL(\Gamma_{\mathsf{L}^p}(M,E),\Gamma_{\mathsf{L}^q}(M,E))$}, 
\]
with the conventions $\left\|\bullet\right\|=\left\|\bullet\right\|_{2}$ and $\left\|\bullet\right\|=\left\|\bullet\right\|_{2,2}$ and analogous notations for functions. The following theorem \ref{xxq} proves the $\mathsf{L}^2 \leadsto\mathsf{L}^q$ smoothing of the Schrödinger semigroup $(\mathrm{e}^{-t H(V)})_{t\geq 0}$. 

\begin{Theorem}\label{xxq} Fix the assumptions of theorem \ref{qp}.\vspace{1.2mm}

{\rm a)} Let $q\in[1,\infty]$ and $t>0$. Then one has the implication
\[
\mathrm{e}^{-t H_0(v)}\in\ILL(\mathsf{L}^2(M),\mathsf{L}^q(M))\>\Rightarrow \>\mathrm{e}^{-t H(V)}\in \ILL\Big(\Gamma_{\mathsf{L}^2}(M,E),\Gamma_{\mathsf{L}^q}(M,E)\Big),
\]
and it holds that
\begin{align}
 \left\| \mathrm{e}^{-t H(V)}\right\|_{2,q}\leq  \left\| \mathrm{e}^{-t H_0(v)}\right\|_{2,q}.\label{djg}
\end {align}
{\rm b)} Assume that  
\begin{align}
C_t:=\sup_{x,y\in M}p_t(x,y)<\infty\>\>\text{ for all $t>0$.}\label{dj7}
\end{align}
Then for any $q\in[2,\infty]$, $t>0$, one has 
\begin{align}
 \mathrm{e}^{-t H(V)}\in\ILL\Big(\Gamma_{\mathsf{L}^2}(M,E),\Gamma_{\mathsf{L}^q}(M,E)\Big),\label{hj1}
\end{align}
in particular, any eigensection of $H(V)$ is in $\Gamma_{\mathsf{L}^q}(M,E)$ for all $q\in[2,\infty]$. More precisely, for any $q\in[2,\infty]$, $t>0$ one has 
\begin{align}
\left\| \mathrm{e}^{-t H(V)}\right\|_{2,q}\leq \sqrt{2} C_t^{\f{1}{2}-\f{1}{q}} \mathrm{e}^{tD(V^{(2)})}, \label{u1}
\end{align}
where $D\left(V^{(2)}\right)> 0$ depends on $V^{(2)}$.
\end{Theorem}

\begin{Remark}{\rm 1. Combining part a) and part b) of theorem \ref{xxq} shows the following surprising fact: If $V$ has a decomposition $V=V^{(1)}-V^{(2)}$ into nonnegative potentials $V^{(j)}$ that satisfy 
\[
\left|V^{(1)}\right|\in\mathsf{L}^1_{\mathrm{loc}}(M)\>\text{ and }\>\left|V^{(2)}\right|\in\mathcal{K}(M),
\]
then the validity of 
\[
 \mathrm{e}^{-t H(V)}\in\ILL\Big(\Gamma_{\mathsf{L}^2}(M,E),\Gamma_{\mathsf{L}^q}(M,E)\Big)
\]
can be achieved by {\it only} requiring additional control on the Riemannian structure of $M$ (namely (\ref{dj7})).\vspace{1.2mm}

2. The inclusion (\ref{hj1}) is contained in proposition 3.5 of \cite{hemp} for scalar operators of the form $H_0(v)$ in the Euclidean $\mathsf{L}^2(\IR^m)$. On the other hand, it seems as if (\ref{hj1}) does not appear in the literature in this form even for operators of the form $H_{\beta}(v)$ in $\mathsf{L}^2(\IR^m)$. In the latter case, however, this result is stated in \cite{Bro1} under the slightly stronger assumption $v^{(1)}\in\mathcal{K}_{\mathrm{loc}}(\IR^m)$, but with $\beta$'s more general than smooth. }
\end{Remark}

We will need the following proposition \ref{as11} for the proof of part b) of theorem \ref{xxq}. Although we will use the result only with $p=2$, it does not cause much extra work to consider the general $\mathsf{L}^p \leadsto\mathsf{L}^q$ situation:

\begin{Proposition}\label{as11} Assume that  
\begin{align}
C_t:=\sup_{x,y\in M} p_t(x,y) <\infty\>\>\text{ for all $t>0$.} 
\end{align}
Then the assignment\footnote{Note that $P_th=\mathrm{e}^{\f{t}{2}\Delta}h$ for $h\in\mathsf{L}^2(M)$.}
\[
P_th(x):=\int_M p_t(x,y) h(y)\mathrm{vol}(\Id y)
\]
defines an element of $\ILL(\mathsf{L}^p(M),\mathsf{L}^q(M))$ for all $1\leq p\leq q\leq \infty$, $t>0$, and one has 
\begin{align}
\left\|P_t\right\|_{p,q}\leq C_t^{\f{1}{p}-\f{1}{q}}.\label{xx2}
\end{align}
\end{Proposition}

The proof of proposition \ref{as11} will be given in section \ref{th}. A short look at the proof shows that proposition \ref{as11} actually has a natural generalization to symmetric essentially bounded integral kernels on $\sigma$-finite measure spaces.\\
Now we can prove theorem \ref{xxq}:\vspace{1.1mm}

{\it Proof of theorem \ref{xxq}.} a) This follows from (\ref{api}). \vspace{1.2mm}

b) We will use part a) for the proof: Setting 
\[
v:=\underline{V}=\min\sigma(V)=\min\sigma(V^{(1)})-\max\sigma(V^{(2)})=: v^{(1)}-v^{(2)}, 
\]
it is sufficient to show that
\[
\mathrm{e}^{-t H_0(v)}\in\ILL(\mathsf{L}^2(M),\mathsf{L}^q(M)).
\]
Note that using $-v\leq v^{(2)}$, proposition \ref{xdd} implies   
\begin{align}
\underset{x\in M}{\mathrm{ess \ sup}}\> \mathbb{E}\left[ \mathrm{e}^{-\beta\int^t_0 v(B_s(x)) \Id s}1_{\{t<\zeta(x)\}}\right] \leq 2  \mathrm{e}^{t C(\beta v^{(2)})}\>\>\text{ for all $\beta\geq 0$}.\label{dxx}
\end{align}

Throughout, let $h\in \mathsf{L}^2(M)$. \vspace{1mm}

Case $q=\infty$: One has
\begin{align}
\left\|\mathrm{e}^{-t H_0(v)}h\right\|_{\infty}&\leq \underset{x\in M}{\mathrm{ess \ sup}} \>\mathbb{E} \left[  \mathrm{e}^{-\int^t_0 v(B_s(x)) \Id s}  | h(B_t(x))| 1_{\{t<\zeta(x)\}}\right]\nn\\
& \leq \underset{x\in M}{\mathrm{ess \ sup}} \>\mathbb{E} \left[  \mathrm{e}^{-2\int^t_0 v(B_s(x)) \Id s}  1_{\{t<\zeta(x)\}}\right]^{\f{1}{2}}\nn\\
&\>\>\>\>\times \underset{x\in M}{\mathrm{ess \ sup}}\>\mathbb{E} \left[  | h(B_t(x))|^{2} 1_{\{t<\zeta(x)\}}\right]^{\f{1}{2}}\nn\\
& \leq   \left(2 \mathrm{e}^{t C\left(2v^{(2)}\right)}\right)^{\f{1}{2}}\left\|\mathrm{e}^{\f{t}{2}\Delta}|h|^2\right\|_{\infty}\nn\\
&\leq \left(2C_t  \mathrm{e}^{t C\left(2v^{(2)}\right)} \right)^{\f{1}{2}}         \left\|h\right\|_2,
\end{align}
where we have used Cauchy-Schwarz for the second step and (\ref{xx2}) for the last step.\vspace{1mm}

Case $q<\infty$: We set $l:=q/2$. Then
\begin{align}
\left\|\mathrm{e}^{-t H_0(v)}h\right\|^q_q&\leq \int_M \mathbb{E} \left[  \mathrm{e}^{-\int^t_0 v(B_s(x)) \Id s} |h(B_t(x))| 1_{\{t<\zeta(x)\}}\right]^q\mathrm{vol}(\Id x)\nn\\
& \leq \int_M   \mathbb{E} \left[  \mathrm{e}^{-2\int^t_0 v(B_s(x)) \Id s}1_{\{t<\zeta(x)\}}\right]^{l}\mathbb{E} \left[  | h(B_t(x))|^{2}1_{\{t<\zeta(x)\}} \right]^{l} \mathrm{vol}(\Id x)\nn\\
& \leq 2^l\mathrm{e}^{lt C\left(2 v^{(2)}\right)} \int_M   \mathbb{E} \left[  | h(B_t(x))|^{2} 1_{\{t<\zeta(x)\}}\right]^{l} \mathrm{vol}(\Id x)\nn
\end{align}
follows again from Cauchy-Schwarz. Finally, we get from (\ref{xx2}) the inequalities
\begin{align}
\left\|\mathrm{e}^{-t H_0(v)}h\right\|_q&\leq \left(2 \mathrm{e}^{t C\left(2 v^{(2)}\right)}\right)^{\f{1}{2}} \left\|\mathrm{e}^{\f{t}{2}\Delta}\right\|^{\f{l}{q}}_{1,l}\left\||h|^2\right\|^{\f{l}{q}}_1 = \left(2\mathrm{e}^{t C\left(2 v^{(2)}\right)}\right)^{\f{1}{2}} \left\|\mathrm{e}^{\f{t}{2}\Delta}\right\|^{\f{l}{q}}_{1,l}\left\|h\right\|_2\nn\\
&\leq \left(2\mathrm{e}^{t C\left(2 v^{(2)}\right)}\right)^{\f{1}{2}} C_t^{\f{1}{2}-\f{1}{q}} \left\|h\right\|_2.
\end{align}
This completes the proof.  \vspace{0.5mm}

\hfill$\blacksquare$\vspace{1.1mm}

The assumption (\ref{dj7}) is satisfied for all $t>0$ in the situation of theorem \ref{Aa}:

\begin{Proposition}\label{hyp} Under the assumptions of theorem \ref{Aa}, for any $t>0$ there are $c_t,d_t>0$, which depend on the Riemannian structure of $M$, such that for all $0<s\leq t$ and all $x,y\in M$ one has
\begin{align}
p_s(x,y)\leq \f{c_t \mathrm{e}^{-d_t \f{\Id(x,y)^2}{s}}}{s^{m/2}}.\label{hyp2}
\end{align}
In particular,  
\begin{align}
C_t:= \sup_{x,y\in M} p_t(x,y)<\infty \>\>\text{ for all $t>0$.}\label{hyp1}
\end{align}
\end{Proposition}

{\it Proof.} The estimate (\ref{hyp2}) follow easily from the considerations of p.110 in \cite{kt} and a simple rescaling argument for the Riemannian structure of $M$. \vspace{0.5mm}

\hfill$\blacksquare$\vspace{1.1mm}

(\ref{hyp2}) combined with (\ref{u1}) immediately implies:

\begin{Corollary} Under the assumptions of theorem \ref{Aa}, let $V$ be such that there is a decomposition $V=V^{(1)}-V^{(2)}$ into potentials $V^{(1)},V^{(2)}\geq 0$ with  
\[
\left|V^{(1)}\right|\in\mathsf{L}^1_{\mathrm{loc}}(M)\>\text{ and }\>\left|V^{(2)}\right|\in\mathcal{K}(M).
\]
Then for all $q\in[2,\infty]$, $0 < t\leq 1$ one has
\begin{align}
\left\| \mathrm{e}^{-t H(V)}\right\|_{2,q}\leq \f{C_q}{t^{\f{m}{2}\left(\f{1}{2}-\f{1}{q}\right)}} \mathrm{e}^{tD(V^{(2)})},
\end{align}
where $C_q>0$ is a constant which only depends on $q$ and the Riemannian structure of $M$ and where $D(V^{(2)})> 0$ depends on $V^{(2)}$.
\end{Corollary}

As a next goal, we want to prove that under very general assumptions on $M$ and $V$ (which should still include practically all physically relevant situations), the operator $\mathrm{e}^{-t H(V)}$ has a
\[
\mathsf{L}^2\leadsto \{\text{bounded continuous}\}
\]
smoothing property for all $t>0$. In detail, this is:

\begin{Theorem}\label{gene} Let $M$ be geodesically complete with Ricci curvature bounded from below and 
\[
\sup_{x,y\in M}p_t(x,y)<\infty\>\>\text{ for all $t>0$.}
\]
Assume furthermore that $V$ is such that there is a decomposition $V=V^{(1)}-V^{(2)}$ into potentials $V^{(1)},V^{(2)}\geq 0$ with  
\begin{align}
\left|V^{(1)}\right|\in\mathcal{K}_{\mathrm{loc}}(M),\>\>\left|V^{(2)}\right|\in\mathcal{K}(M).\label{ssja}
\end{align}
Then for any $t>0$, $f\in\Gamma_{\mathsf{L}^2}(M,E)$, the section 
\begin{align}
M\longrightarrow E,\>\>x\longmapsto  \mathbb{E} \left[  \mathscr{V}^{x}_t  \pa^{x,-1}_t f(B_t(x))\right]\in E_x \label{recht}
\end{align}
is well-defined, continuous and bounded. In particular, $\mathrm{e}^{-tH(V)}f$ has a continuous bounded representative for any $t>0$, $f\in\Gamma_{\mathsf{L}^2}(M,E)$, and each eigensection of $H(V)$ can be chosen continuous and bounded.
\end{Theorem}

Potentials with the property (\ref{ssja}) are usually called {\it Kato decomposable} in the mathematical physics literature.

\begin{Remark}\label{uunv}{\rm 1. As we have already remarked, the above assumptions on the Riemannian structure are satisfied by the class of Riemannian manifolds from theorem \ref{Aa}.\vspace{1mm}

2. Theorem \ref{gene} generalizes theorem 1.9 of \cite{Gue2}, where we have considered the case $V^{(2)}=0$ on Euclidean vector bundles of the form $\IR^m\times\IC^d$. It also generalizes one of the main results of \cite{Bru} (see theorem 21 therein): In the latter paper the authors have considered scalar operators of the form $H_{\beta}(v)$ on $M$'s with a bounded geometry.\vspace{1mm}

3. Let us explain our approach for proving theorem \ref{gene}: In the situation of theorem \ref{gene}, let
\begin{align}
Q^{V}_tf(x):=\mathbb{E} \left[  \mathscr{V}^{x}_t  \pa^{x,-1}_t f(B_t(x))\right].\label{zbbb}
\end{align}
Firstly, we remark that under our assumptions on $M$ and $V$, the right-hand side of (\ref{zbbb}) will indeed turn out to be well-defined for {\it all} $x\in M$ (see proposition \ref{cas}). We will use semigroup domination and theorem \ref{hyp} to prove that $Q^{V}_tf$ is bounded. Furthermore, one can prove that $Q^{V}_{\bullet}f(x)$ satisfies a semigroup property for all $x\in M$ (a priori, this is only clear for {\it a.e.} $x\in M$, and the proof that it remains true for {\it all} $x\in M$ is actually quite technical). From these considerations, it is clear that we may assume
\[
f\in \Gamma_{\mathsf{L}^{\infty}}(M,E)\cap\Gamma_{\mathsf{L}^{2}}(M,E). 
\]
Next, we will use local elliptic regularity to prove that $Q^{0}_t\tilde{f}$ is $\mathsf{C}^{\infty}$ for any $t>0$ and any essentially bounded square integrable $\tilde{f}$, so that the continuity of $Q^{V}_tf$ will follow, if we can locally uniformly approximate $Q^{V}_tf$ as $s\searrow 0$ by $Q^{0}_{s}Q^V_{t-s}f$. This will in fact follow from the perturbation formula (\ref{uhgq}) below and the convergence (\ref{qebb}). The latter of which strongly relies on the assumption that the potential is in the local Kato class. These techniques extend the corresponding ones from \cite{chung} (see also \cite{Bro1}) for usual scalar operators to our setting, where we remark that the proofs of assertions like proposition \ref{kkl3}, proposition \ref{kkl4} or proposition \ref{kkl2} are almost trivial in the setting of \cite{chung}.} 
\end{Remark}

The following five propositions will help us to turn the considerations of remark \ref{uunv}.3 into a full proof. Firstly, we shall prove the asserted well-definedness of the right-hand side of the Feynman-Kac formula. We will actually need a slightly more general result:

\begin{Proposition}\label{cas} Under the assumptions of theorem \ref{gene}, the process $\mathscr{V}^{x}$ exists for all $x\in M$, and for any $0\leq s\leq t$ one has 
\begin{align}
\mathbb{E} \left[ \left\|\mathscr{V}^{x,-1}_s \mathscr{V}^{x}_t  \right\|_x\left\| f(B_t(x))\right\|_x\right]&\leq \left(2\mathrm{e}^{C\left(2\left|V^{(2)}\right|\right)t}\mathrm{e}^{\f{t}{2}\Delta}|f|^2 (x)\right)^{\f{1}{2}}\label{boundl2}\\
&\leq \left(2\mathrm{e}^{C\left(2\left|V^{(2)}\right|\right)t}\sup_{y,z\in M} p_t(y,z)\right)^{\f{1}{2}}\left\|f\right\|.\label{boundedo}
\end{align}
\end{Proposition}

{\it Proof.} Clearly, lemma \ref{Katto0} and the Banach fixed point theorem imply the existence of $\mathscr{V}^{x}$ for all $x$. \\
Noting 
\begin{align}
\left\|\mathscr{V}^{x,-1}_s\mathscr{V}^{x}_t \right\|_x\leq \mathrm{e}^{\int^t_0\left\|V^{(2)}(B_u(x))\right\|_{B_u(x)}\Id u }\>\>\text{ $\mathbb{P}$-a.s.,}
\end{align}
which follows from combining proposition \ref{absc} d) with 
\begin{align}
-V\>\>\leq\>\>-\underline{V}\mathbf{1},\>\>-\underline{V}\>\>\leq\>\> \max\sigma(V^{(2)})\>\> \leq\>\> \left|V^{(2)}\right|,\label{jhgk}
\end{align}
we can use proposition \ref{xdd} to estimate  
\begin{align}
\mathbb{E}\left[\left\|\mathscr{V}^{x,-1}_s\mathscr{V}^{x}_t \right\|^2_x\right]\leq 2 \mathrm{e}^{C\left(2\left|V^{(2)}\right|\right)t}.\label{gjs}
\end{align}

Using Cauchy-Schwarz, (\ref{gjs}) implies
\begin{align}
&\mathbb{E} \left[ \left\|\mathscr{V}^{x,-1}_s \mathscr{V}^{x}_t  \right\|_x \left\|f(B_t(x))\right\|_x\right]\nn\\
&\leq \mathbb{E} \left[ \left\|\mathscr{V}^{x,-1}_s \mathscr{V}^{x}_t\right\|^2_x\right]^{\f{1}{2}}\mathbb{E} \left[ \left\| f(B_t(x))\right\|^2_x\right]^{\f{1}{2}}\nn\\
&=\mathbb{E} \left[ \left\|\mathscr{V}^{x,-1}_s \mathscr{V}^{x}_t\right\|^2_x\right]^{\f{1}{2}}\left(\int_Mp_t(x,y)\left\| f(y)\right\|^2_y\mathrm{vol}(\Id y)\right)^{\f{1}{2}}\nn\\
&\leq \left(2\mathrm{e}^{C\left(2\left|V^{(2)}\right|\right)t}\sup_{y,z\in M} p_t(y,z)\right)^{\f{1}{2}}\left\|f\right\|,
\end{align}
and the proof is complete. \vspace{0.5mm}

\hfill$\blacksquare$\vspace{1.1mm}

Next, we prove the asserted semigroup property and the perturbation formula, respectively:

\begin{Proposition}\label{kkl3} Under the assumptions of theorem \ref{gene}, let  
\begin{align}
Q^{V}_tf(x):=\mathbb{E} \left[  \mathscr{V}^{x}_t  \pa^{x,-1}_t f(B_t(x))\right]\>\>\text{ for any $t\geq 0$, $x\in M$.}\label{uhgq0}
\end{align}
{\rm a)} $Q^V_{\bullet}f$ satisfies a pointwise semigroup identity, 
\begin{align}
Q^V_{s+t}f(x)=Q^V_sQ^V_tf(x)\>\>\text{ for any $s,t\geq 0$, $x\in M$.}\label{uhgq1}
\end{align}

{\rm b)} One has the following perturbation formula for any $t\geq s\geq  0$, $x\in M$,  
\begin{align}
Q^0_sQ^V_{t-s}f(x)=\mathbb{E} \left[  \mathscr{V}^{x,-1}_s\mathscr{V}^{x}_t  \pa^{x,-1}_t f(B_t(x))\right].\label{uhgq}
\end{align}
\end{Proposition}

{\it Proof.} Note first that all terms in (\ref{uhgq0}) - (\ref{uhgq}) are indeed pointwise well-defined, which is implied by proposition \ref{cas} and proposition \ref{as11}. The proposition will now be proved in four steps. \vspace{1.2mm}

I) {\it (\ref{uhgq1}) and (\ref{uhgq}) hold under the additional assumptions $|V|\in\mathsf{L}^{\infty}(M)$ and $f\in\Gamma_{\mathsf{L}^{\infty}}(M,E)$.} \vspace{1.2mm}

Proof: We have to introduce some notation first: In view of (\ref{dk2j}), for any starting time $a\geq 0$ and any appropiate $\IFF_{a}$-measurable $h:\Omega\to M$, we define the processes $B^{a,h}$, $\pa^{a,h}$ and $\mathscr{V}^{a,h}$ as follows: 
\[
 B^{a,h}:[a,\infty)\times \Omega\longrightarrow M
\]
is defined as the maximal solution of 
\[
\Id B^{a,h}=\sum^l_{j=1} A_j(B^{a,h}) \underline{\Id} W^j,\>\>B^{a,h}_{a}=h,
\]
$\pa^{a,h}$ is defined as the stochastic parallel transport corresponding to $B^{a,h}$, so that 
\[
\pa^{a,h}_b:E_{h}\longrightarrow E_{ B^{a,h}_b}\>\>\text{ for any $b\geq a$},
\]
and, finally, for $\mathbb{P}$-a.e. $\omega\in\Omega$, the map 
\[
\mathscr{V}^{a,h}_{\bullet}(\omega):[0,\infty)\longrightarrow \mathrm{End}(E)_{h(\omega)}
\]
is defined as the weak solution of 
\begin{align}
\Id  \mathscr{V}^{a,h}_{t}(\omega) = -\mathscr{V}^{a,h}_{t}(\omega)\Big( \pa^{a,h,-1}_{a+t} V(B^{a,h}_{a+t}) \pa^{a,h}_{a+t}\Big)(\omega)\Id t,\>\mathscr{V}^{a,h}_{0}(\omega)=\mathbf{1}.\nn
\end{align} 
Note that $\mathscr{V}^{a,h}_{t}(\omega)$ can be expanded as in (\ref{pato}) and that our usual notation implies
\[
(B^{0,x}, \pa^{0,x},\mathscr{V}^{0,x})=(B(x), \pa^{x},\mathscr{V}^{x}).
\]

Proof of (\ref{uhgq1}): Let $U^x$ be a lift of $B(x)$ and let $U^{s,B_s(x)}$ be the lift of $B^{s,B_s(x)}$ from $U^x_s$. Then we have $\pa^{x}=U^xU^{x,-1}_0$ and $\pa^{s,B_s(x)}=U^{s,B_s(x)}U^{x,-1}_s$, so that the flow property of the solutions of
\begin{align}
\Id U=\sum^l_{j=1} A^*_j(U) \underline{\Id} W^j\label{b2}
\end{align}
gives
\begin{align}
\pa^{x}_{s+t}=\pa^{s,B_s(x)}_{s+t}\pa^{x}_s\>\>\text{ $\mathbb{P}$-a.s.}\label{uun}
\end{align}
Using (\ref{uun}) and the flow property of the solutions of
\begin{align}
\Id B=\sum^l_{j=1} A_j(B) \underline{\Id} W^j,\label{b1}
\end{align}
one easily checks that for fixed $s$, the processes 
\[
 \mathscr{V}^{x}_{s+\bullet} \>\>\text{ and }\>\>\mathscr{V}^{x}_s\pa^{x,-1}_{s}\mathscr{V}^{s,B_s(x)}_{\bullet}\pa^{x}_{s} 
\]
both solve the same $\mathrm{End}(E)_x$-valued initial value problem, so that by uniqueness and (\ref{uun}) we get the multiplicative property
\begin{align}
\mathscr{V}^{x}_{s+t}\pa^{x,-1}_{s+t}=\mathscr{V}^{x}_s\pa^{x,-1}_{s}\mathscr{V}^{s,B_s(x)}_{t}\pa^{s,B_s(x),-1}_{s+t}\>\>\text{ $\mathbb{P}$-a.s.}\label{a6}
\end{align}
With $\mathbb{E}^{\IFF_s}[\bullet]:=\mathbb{E}[\bullet|\IFF_s]$ the last identity implies
\begin{align}
Q^V_{s+t}f(x)&=\mathbb{E}\left[\mathscr{V}^{x}_s\pa^{x,-1}_{s}\mathscr{V}^{s,B_s(x)}_{t}\pa^{s,B_s(x),-1}_{s+t}f(B_{s+t}(x)) \right]\nn\\
&=\mathbb{E}\left[\mathscr{V}^{x}_s\pa^{x,-1}_{s}\mathbb{E}^{\IFF_s}\left[\mathscr{V}^{s,B_s(x)}_{t}\pa^{s,B_s(x),-1}_{s+t}f\left( B^{s,B_s(x)}_{s+t}\right)  \right]\right].
\end{align}
Since $\IFF_s$ is independent from $\IFF_{s+t}$ and since by its definition $\mathscr{V}^{s,B_s(x)}_{t}$ clearly is an $\IFF_{s+t}$ - random variable, we can use lemma 6.3.1 in \cite{Ha} to conclude
\begin{align}
&\mathbb{E}\left[\mathscr{V}^{x}_s\pa^{x,-1}_{s}\mathbb{E}^{\IFF_s}\left[\mathscr{V}^{s,B_s(x)}_{t}\pa^{s,B_s(x),-1}_{s+t}f\left( B^{s,B_s(x)}_{s+t}\right)  \right]\right]\nn\\
=& \int_{\Omega}\mathscr{V}^{x}_s(\omega)\pa^{x,-1}_{s}(\omega)\int_{\Omega}Z^{s,B_s(x)(\omega)}_{t}(\tilde{\omega})\mathbb{P}(\Id \tilde{\omega}) \ \mathbb{P}(\Id \omega),
\end{align}
with
\[
Z^{a,y}_{t}:=\mathscr{V}^{a,y}_{t}\pa^{a,y}_{a+t}f\left( B^{a,y}_{a+t}\right)\>\>\text{for any $a\geq 0$, $y\in M$.}
\]
It follows that it is sufficient to prove 
\begin{align}
\mathbb{E}\left[Z^{s,y}_{t}\right]=\mathbb{E}\left[Z^{0,y}_{t}\right]\>\>\text{ for any $y\in M$}.\label{a3}
\end{align}
Let $\pi:\mathrm{P}(E)\to M$ denote the principal bundle projection, let $U^y$ be a lift of $B(y)$ and let $U^{s,y}$ be the lift of $B^{s,y}$ from $U^y_0$. Since parallel transport does not depend on the particular choice of the initial frame, we have $\pa^{s,y}= U^{s,y}U^{y,-1}_0$, and clearly we have $B^{s,y}=\pi(U^{s,y})$, $\pa^{y}= U^yU^{y,-1}_0$, $B(y)=\pi(U^y)$. For any $y\in M$ and $n\in \IN$ we define a function $\IAA^{t,y}_n$ by setting
\begin{align}
&\IAA^{t,y}_n :\mathsf{C}([0,\infty),\mathrm{P}(E))\longrightarrow E_y,\nn\\
&\IAA^{t,y}_n(\gamma[\bullet]):=\left\lbrace  \prod^{\longrightarrow}_{1\leq j\leq n} 
\left(\mathbf{1}+ \f{t}{n} U^y_0 \gamma[(tj)/n]^{-1} V\Big(\pi\left( \gamma[ (tj)/n]\right)\Big)\gamma[(tj)/n] U^{y,-1}_0 \right) \right\rbrace \nn\\
&\>\>\>\>\>\>\>\>\>\>\>\>\>\>\>\times U^{y}_0 \gamma[ t]^{-1} f\Big( \pi\left( \gamma[ t]\right) \Big) .
\end{align}
Then we have the following inequalities, 
\begin{align}
\left\| \IAA^{t,y}_n(U^{s,y}_{s+\bullet})\right\|_{y}\leq \mathrm{e}^{t \left\| V\right\|_{\infty}} \left\| f(B^{s,y}_{s+t})\right\|_{B^{s,y}_{s+t}}\>\>\text{ $\mathbb{P}$-a.s.}\label{a1}
\end{align}
and
\begin{align}
\left\| \IAA^{t,y}_n(U^{y})\right\|_{y}\leq \mathrm{e}^{t \left\| V\right\|_{\infty}} \left\| f(B_t(y))\right\|_{B_t(y)}\>\>\text{ $\mathbb{P}$-a.s.}\label{a2}
\end{align}
Since $\mathscr{V}^{s,y}$ and $\mathscr{V}^{(y)}$ can be represented as product integrals (this follows from applying theorem 7.1 in \cite{dollard} with $z\mapsto 1+z$ together with the corresponding remarks on page 56), one has
\[
 \lim_{n\to\infty}\mathscr{A}^{t,y}_{n}(U^{s,y}_{s+\bullet})=Z^{s,y}_{t}\>\>\text{ and }\>\> \lim_{n\to\infty}\mathscr{A}^{t,y}_{n}(U^{y})=Z^{0,y}_{t}\>\>\text{ $\mathbb{P}$-a.s.}
\]
Now we note that $U^{s,y}_{s+\bullet}$ and $U^{y}$ have the same law\footnote{To see this, note first that the smoothness of the vector fields $A^*_j$ implies the uniqueness in law for (\ref{b2}) (this follows from theorem 1.1.10 in \cite{Hsu} and the Whitney embedding theorem). Now one can use the same arguments as in the proof of corollary 1 to Satz 6.40 in \cite{Ha} to deduce that $U^{s,y}_{s+\bullet}$ and $U^{y}$ are equal in law.}, so that we can use dominated convergence (in view of (\ref{a1}) and (\ref{a2})) to deduce
\[
\mathbb{E}\left[Z^{s,y}_{t}\right]= \lim_{n\to\infty}\mathbb{E}\left[\mathscr{A}^{t,y}_{n}(U^{s,y}_{s+\bullet})\right] = \lim_{n\to\infty}\mathbb{E}\left[\mathscr{A}^{t,y}_{n}(U^{y})\right] =\mathbb{E}\left[Z^{0,y}_{t}\right].
\]

Proof of (\ref{uhgq}): We calculate
\begin{align}
Q^0_sQ^V_{t-s}f(x)&=\int_{\Omega} \pa^{x,-1}_s(\omega) \int_{\Omega} \mathscr{V}^{(B_s(x)(\omega))}_{t-s}(\tilde{\omega})\pa^{B_s(x)(\omega)}_{t-s}(\tilde{\omega}) \nn\\
&\>\>\>\>\>\>\>\>\>\>\>\>\>\times f\left(    B_{t-s}\Big(  B_s(x)(\omega) \Big)  (\tilde{\omega})     \right)   \mathbb{P}(\Id \tilde{\omega} ) \ \mathbb{P}(\Id \omega)\nn\\
&=\int_{\Omega} \pa^{x,-1}_s(\omega) \int_{\Omega} \mathscr{V}^{s,B_s(x)(\omega)}_{t-s}(\tilde{\omega})\pa^{s,B_s(x)(\omega)}_{t}(\tilde{\omega}) \nn\\
&\>\>\>\>\>\>\>\>\>\>\>\>\>\times f\left(    B^{s, B_s(x)(\omega)}_{t}(\tilde{\omega})  \right)      \mathbb{P}(\Id \tilde{\omega} ) \ \mathbb{P}(\Id \omega)\nn\\
&=\mathbb{E}\left[\pa^{x,-1}_{s}\mathbb{E}^{\IFF_s}\left[\pa^{x}_{s}\mathscr{V}^{x,-1}_{s}\mathscr{V}^{x}_{t}\pa^{x,-1}_{t}f\left( B^{s,B_s(x)}_{t}\right)  \right]\right]\nn\\
&=\mathbb{E}\left[\mathscr{V}^{x,-1}_{s}\mathscr{V}^{x}_{t}\pa^{x,-1}_{t}f\left( B_{t}(x)\right)\right],
\end{align}
where we have used (\ref{a3}) for the second equality, lemma 6.3.1 in \cite{Ha} together with (\ref{a6}) for the third equality, and the flow property of (\ref{b1}) for the last equality. \vspace{1.2mm}

II) {\it (\ref{uhgq1}) and (\ref{uhgq}) hold under the additional assumptions $V\geq C$ and $f\in\Gamma_{\mathsf{L}^{\infty}}(M,E)$.}\vspace{1.2mm}

Proof of (\ref{uhgq1}): We can assume $V\geq 0$ and we define $V_n:=\min(n,V)$ for any $n\in\IN$. Then each $V_n$ is a bounded potential, so that by applying II) implies that for all $n$,
\begin{align}
Q^{V_n}_{s+t}f(x)=Q^{V_n}_sQ^{V_n}_tf(x).\label{l44}
\end{align}
Furthermore, the following two identities are included in the first part of the proof of theorem \ref{hb3}: For all $a\geq 0$, $y\in M$, 
\begin{align}
\lim_{n\to\infty} \mathscr{V}^{(y)}_{n,a}=\mathscr{V}^{(y)}_{a}\>\>\text{ $\mathbb{P}$-a.s.}\label{y2}
\end{align}
(with an obvious notation), and
\begin{align}
\lim_{n\to\infty} Q^{V_n}_a f(y)= Q^{V}_a f(y).\label{y1}
\end{align}
Thus it remains to prove 
\begin{align}
\lim_{n\to\infty}\mathbb{E}\left[\mathscr{V}^{x}_{n,s}  \pa^{x,-1}_s Q^{V_n}_tf(B_s(x))\right]=\mathbb{E}\left[\lim_{n\to\infty}\mathscr{V}^{x}_{n,s}  \pa^{x,-1}_s Q^{V_n}_tf(B_s(x))\right].\label{x6}
\end{align}
To this end, we remark 
\begin{align}
\left\|\mathscr{V}^{(y)}_{n,a}\right\|_y\leq 1\>\>\text{$\mathbb{P}$-a.s. for all $y\in M$, $n\in \IN$, $a\geq 0$}\label{z5}
\end{align}
(this is also included in the first part of the proof of theorem \ref{hb3}), so that
\begin{align}
\left\|\mathscr{V}^{x}_{n,s}  \pa^{x,-1}_s Q^{V_n}_tf(B_s(x))\right\|_x \leq \left\|f\right\|_{\infty}\>\>\text{ $\mathbb{P}$-a.s.,}
\end{align}
and (\ref{x6}) follows from dominated convergence. \vspace{1.2mm}

Proof of (\ref{uhgq}): Again, we may assume $V\geq 0$ and we define $V_n:=\min(n,V)$. Then by II) we have for all $n$,
\begin{align}
Q^0_sQ^{V_n}_{t-s}f(x)=\mathbb{E} \left[  \mathscr{V}^{x,-1}_{n,s}\mathscr{V}^{x}_{n,t}  \pa^{x,-1}_t f(B_t(x))\right]. 
\end{align}
Furthermore, (\ref{y2}) implies 
\begin{align}
\lim_{n\to\infty} \mathscr{V}^{x}_{n,t}=\mathscr{V}^{x}_{t},\>\lim_{n\to\infty} \mathscr{V}^{x,-1}_{n,s}=\mathscr{V}^{x,-1}_{s}\>\>\text{ $\mathbb{P}$-a.s.,}\label{x2}
\end{align} 
and we can use 
\begin{align}
\left\|\mathscr{V}^{x,-1}_{n,s}\mathscr{V}^{x}_{n,t}  \pa^{x,-1}_t f(B_t(x))\right\|_x\leq \left\|f\right\|_{\infty}\>\>\text{$\mathbb{P}$-a.s.}
\end{align}
(which follows from $-\underline{V_n}\leq 0$ and proposition \ref{absc} d)) to conclude 
\begin{align}
\lim_{n\to\infty}\mathbb{E} \left[  \mathscr{V}^{x,-1}_{n,s}\mathscr{V}^{x}_{n,t}  \pa^{x,-1}_t f(B_t(x))\right]=\mathbb{E} \left[  \mathscr{V}^{x,-1}_{s}\mathscr{V}^{x}_{t}  \pa^{x,-1}_t f(B_t(x))\right].\label{x3}
\end{align}
It remains to prove 
\begin{align}
\lim_{n\to\infty}Q^0_sQ^{V_n}_{t-s}f(x)=Q^0_sQ^{V}_{t-s}f(x).\label{f7}
\end{align}
To this end, we just note that by (\ref{z5}) we have
\begin{align}
\left\|\pa^{x,-1}_s Q^{V_n}_{t-s}f(B_s(x))\right\|_x\leq \left\|f\right\|_{\infty}\>\>\text{$\mathbb{P}$-a.s.,}
\end{align}
so that (\ref{f7}) follows from dominated convergence and (\ref{y1}). \vspace{1.2mm}

III) {\it (\ref{uhgq1}) and (\ref{uhgq}) hold under the additional assumption $f\in\Gamma_{\mathsf{L}^{\infty}}(M,E)$.}\vspace{1.2mm}

Proof of (\ref{uhgq1}): We define $V_n:=\max(-n,V)$ for any $n\in \IN$. Then each $V_n$ is a bounded from below, locally Kato potential. By II) we have (\ref{l44}) and it follows from the second part of the proof of theorem \ref{hb3} that one also has (\ref{y2}) again. For the proof of (\ref{y1}) note that for all $y\in M$, $n\in \IN$, $a\geq 0$ one has
\begin{align}
\left\|\mathscr{V}^{(y)}_{n,a}\right\|_y\leq \mathrm{e}^{-\int^a_0 \underline{V}(B_u(y))\Id u} \leq \mathrm{e}^{\int^a_0 \left\|V^{(2)}(B_u(y))\right\|_{B_u(y)}\Id u}\label{y5}
\end{align}
(only the first inequality nontrivial; but this is included in the second part of the proof of theorem \ref{hb3}), so
\begin{align}
\left\|\mathscr{V}^{(y)}_{n,a}  \pa^{x,-1}_a f(B_a(y))\right\|_x\leq \left\|f\right\|_{\infty}\mathrm{e}^{\int^a_0 \left\|V^{(2)}(B_u(x))\right\|_{B_u(x)}\Id u}\>\>\text{ $\mathbb{P}$-a.s.}
\end{align}
and the last term is in $\mathsf{L}^1(\mathbb{P})$ by proposition \ref{xdd}, so that (\ref{y1}) follows from (\ref{y2}) and dominated convergence. It remains to prove (\ref{x6}). But in view of (\ref{y5}) we have
\begin{align}
&\left\|\mathscr{V}^{x}_{n,s}  \pa^{x,-1}_s Q^{V_n}_tf(B_s(x))\right\|_x \nn\\
&\leq \left\|f\right\|_{\infty}\mathrm{e}^{\int^s_0 \left\|V^{(2)}(B_u(x))\right\|_{B_u(x)}\Id u}\ \mathbb{E}\left\| \mathscr{V}^{(y)}_{n,t}\right\|_y\mid_{y=B_s(x)} \nn\\
&\leq \left\|f\right\|_{\infty}\mathrm{e}^{\int^s_0 \left\|V^{(2)}(B_u(x))\right\|_{B_u(x)}\Id u} \ \mathbb{E}\left[\mathrm{e}^{\int^t_0 \left\|V^{(2)}(B_u(y))\right\|_{B_u(y)}\Id u}\right]\mid_{y=B_s(x)} \nn\\
&\leq 2\left\|f\right\|_{\infty}\mathrm{e}^{ t C\left(\left|V^{(2)}\right|\right)}\mathrm{e}^{\int^s_0 \left\|V^{(2)}(B_u(x))\right\|_{B_u(x)}\Id u}\>\>\text{ $\mathbb{P}$-a.s.},
\end{align}
and the last term is in $\mathsf{L}^1(\mathbb{P})$ by proposition \ref{xdd}. Now (\ref{x6}) follows from dominated convergence.\vspace{1.2mm}

Proof of (\ref{uhgq}): Again, let $V_n$ be given by $\max(-n,V)$. Then by (\ref{y2}) we have (\ref{x2}), and one furthermore has 
\begin{align}
\left\|\mathscr{V}^{x,-1}_{n,s}\mathscr{V}^{x}_{n,t}  \pa^{x,-1}_t f(B_t(x))\right\|_x\leq \left\|f\right\|_{\infty} \mathrm{e}^{\int^t_0\left\|V^{(2)}(B_u(x))\right\|_{B_u(x)}}\in \mathsf{L}^1(\mathbb{P})
\end{align}
(which follows from proposition \ref{absc} d) and $-\underline{V_n}\leq -\underline{V}$; the latter inequality is included in the second part of the proof of theorem \ref{hb3}) so that we have (\ref{x3}) by dominated convergence. It remains to prove (\ref{f7}). To this end, we can use (\ref{y1}) and
\begin{align}
&\left\|\pa^{x,-1}_s Q^{V_n}_{t-s}f(B_s(x))\right\|_x\leq \left\|f\right\|_{\infty}\mathbb{E}\left[\left\|\mathscr{V}^{(y)}_{n,t-s}\right\|_y\right]\mid_{y=B_s(x)}\nn\\
&\leq  \left\|f\right\|_{\infty}  \mathbb{E}\left[\mathrm{e}^{\int^{t-s}_0 \left\|V^{(2)}(B_u(y))\right\|_{B_u(y)}\Id u}\right]\mid_{y=B_s(x)}\nn\\
& \leq 2\left\|f\right\|_{\infty}  \mathrm{e}^{(t-s)C\left(\left|V^{(2)}\right|\right)}\>\>\text{ $\mathbb{P}$-a.s.,}
\end{align}
which follows from (\ref{y5}) and proposition \ref{xdd}, to deduce (\ref{f7}) with dominated convergence again.

IV) {\it (\ref{uhgq1}) and (\ref{uhgq}) hold in the general situation.}\vspace{1.2mm}

Proof: It remains to remove the condition that $f$ is bounded. To this end, one can consider $f_n:=1_{\mathrm{K}_n(\mathscr{O})}f\in\Gamma_{\mathsf{L}^{\infty}}(M,E)$ for some fixed reference point $\mathscr{O}$, apply III) to the $f_n$'s and take $n\to\infty$ to deduce this assertion with dominated convergence. \vspace{0.5mm}

\hfill$\blacksquare$\vspace{1.1mm}

Next, we will prove:

\begin{Proposition}\label{kkl4} Let $M$ be stochastically complete. Then for any $t>0$ and any $f\in \Gamma_{\mathsf{L}^{\infty}}(M,E)\cap\Gamma_{\mathsf{L}^{2}}(M,E)$, the section given by
\begin{align}
M\longrightarrow E,\>\>x\longmapsto \mathbb{E} \left[   \pa^{x,-1}_t f(B_t(x))\right]\in E_x
\end{align}
is the $\mathsf{C}^{\infty}$-representative of $\mathrm{e}^{-t H(0)}f$. 
\end{Proposition}

{\it Proof.} By local elliptic regularity, $\mathrm{e}^{-sH(0)}f$ has a $\mathsf{C}^{\infty}$-representative which we denote with $f_s(\bullet)$ for all $s>0$. Furthermore, the map $(s,y)\mapsto f_s(y)$ is $\mathsf{C}^{\infty}$ and one has 
\begin{align}
\partial_s f_s(y)=-\f{1}{2} \nabla^*\nabla f_s(y)\>\>\text{ for all $s>0$, $y\in M$.}\label{nn33}
\end{align}
We fix arbitrary $x\in M$, $t>0$ and $\epsilon>0$ now. Then the time dependent version of formula (12) from \cite{Gue} combined with the above (\ref{nn33}) gives
\begin{align}
&\Id_s \Big(\pa^{x,-1}_sf_{t-s+\epsilon}(B_s(x))\Big)\nn\\
&= \pa^{x,-1}_s \sum^{l}_{j=1}(\nabla_{A_j}f_{t-s+\epsilon})(B_s(x)) \Id W^j_s  -\f{1}{2}\pa^{x,-1}_s\nabla^*\nabla f_{t-s+\epsilon}(B_s(x)) \Id s\nn\\
&\>\>\>\>\>\>\>\>+\pa^{x,-1}_s\partial_s f_{t-s+\epsilon}(B_s(x))\Id s\nn\\
&= \pa^{x,-1}_s \sum^{l}_{j=1}(\nabla_{A_j}f_{t-s+\epsilon})(B_s(x)) \Id W^j_s \label{ssu}
\end{align}
for all $0\leq s\leq t$, which implies that the process
\[
N(x,t,\epsilon):[0,t]\times\Omega\longrightarrow E_x,\>\>N_s(x,t,\epsilon):=\pa^{x,-1}_sf_{t-s+\epsilon}(B_s(x)) \label{gelb}
\]
is a continuous local martingale. It is in fact a martingale: For any $0\leq s\leq  t$ the following inequalities hold $\mathbb{P}$-a.s.,
\begin{align}
 \left\|N_s(x,t,\epsilon)\right\|_x&\leq  \left\|f_{t-s+\epsilon}(B_s(x))\right\|_{B_s(x)}\nn\\
&\leq   \int_M p_{t-s+\epsilon}(B_s(x),y) \left\|f(y)\right\|_y\mathrm{vol}(\Id y)\nn\\
&\leq  \left\|f\right\|_{\infty} \int_M p_{t-s+\epsilon}(B_s(x),y) \mathrm{vol}(\Id y)=  \left\|f\right\|_{\infty},\label{ak74}
\end{align}
where we have used (\ref{api}) with $V=0$, $v=0$ for the second inequality\footnote{Note that (\ref{api}) is true for {\it all $x$} in the $\mathsf{C}^{\infty}$ case.}, so that the martingale property of $N(x,t,\epsilon)$ follows from a standard criterion (see for example p.129 in \cite{Rev}). This shows 
\begin{align}
f_{t+\epsilon}(x)= \mathbb{E}\left[N_0(x,t,\epsilon)\right]= \mathbb{E}\left[ N_t(x,t,\epsilon)\right]=\mathbb{E}\left[\pa^{x,-1}_tf_{\epsilon}(B_t(x)) \right].\label{ghe}
\end{align}
Since $f_s=\mathrm{e}^{-sH(0)}f\to f$ as $s \searrow 0$ with respect to $\Gamma_{\mathsf{L}^2}(M,E)$, there exists a sequence $(\epsilon_n)\subset (0,\infty)$ with $\epsilon_n\to 0$ and $f_{\epsilon_n}(y)\to f(y)$ as $n\to\infty$ for a.e. $y\in M$, so that (in view of (\ref{ak74})) we can use dominated convergence and (\ref{ghe}) with $\epsilon=\epsilon_n$ to conclude
\begin{align}
\mathbb{E}\left[\pa^{x,-1}_tf(B_t(x)) \right]= \lim_{n\to\infty}\mathbb{E}\left[\pa^{x,-1}_tf_{\epsilon_n}(B_t(x)) \right]=\lim_{n\to\infty} f_{t+\epsilon_n}(x) =f_{t}(x),\nn
\end{align}
and the proof is complete. \vspace{0.5mm}

\hfill$\blacksquare$\vspace{1.1mm}

The next proposition is concerned with the first exit time of $B(x)$ from geodesic balls, where $x$ runs through a compact set. Although the arguments of the proof that we are going to present are certainly well-known from proofs of stochastic completeness, the result itself has not yet appeared in the literature, as far as we know.

\begin{Proposition}\label{kkl} Let $M$ be geodesically complete with Ricci curvature bounded from below, fix some origin $\mathscr{O}\in M$, and for any $t>0$, $x\in M$, $r>0$ let $\chi(r,t,x):= 1_{ \left\lbrace t<\zeta(r,x)\right\rbrace}$, where $\zeta(r,x)$ stands for the first exit time of $B(x)$ from $\mathrm{K}_r(\mathscr{O})$. If $K\subset M$ is compact, then one has 
\begin{align}
 \lim_{t \searrow 0} \sup_{x\in K} \mathbb{E}\left[1-\chi(r,t,x) \right]=0\>\>\text{ for any $r> \max_{x\in K} \Id(\mathscr{O},x)$.} \label{in}
\end{align}
\end {Proposition}

\begin{Remark}{\rm 1. Note that (\ref{in}) is nothing but 
\begin{align}
 \lim_{t\searrow 0} \inf_{x\in K} \mathbb{P}  \{t < \zeta(r,x)\} =1 \>\>\text{ for any $r> \max_{x\in K} \Id(\mathscr{O},x)$.}\label{ggd1}
\end{align}


2. By using the techniques of \cite{Hsu2}, it should be possible to relax the assumption on the Ricci curvature considerably. }

\end{Remark}

{\it Proof of proposition \ref{kkl}.} Since $B(x)$ is continuous, we can assume $K\ne\{\mathscr{O}\}$. Let $R(x):=\Id(\mathscr{O},x)$. Then $R$ is a smooth function on the open set $M\setminus (\mathrm{Cut}(\mathscr{O})\cup\{\mathscr{O}\})$. If $C>0$ is such that the Ricci curvature of $M$ is bounded from below by $-C$, then inequality (2.3) of \cite{qian} implies 
\begin{align}
\Delta R(x) \leq  h(R(x))\>\>\text{ for all $x\in M\setminus (\mathrm{Cut}(\mathscr{O})\cup\{\mathscr{O}\})$},\label{ecm2}
\end{align}
where 
\[
h:(0,\infty)\longrightarrow (0,\infty),\>\>h(r):=\f{m-1}{r }+ \f{C}{3} r.
\]
Furthermore, although $R \notin\mathsf{C}^{\infty}(M)$, the process $R(B(x))$ is a continuous semi-martingale \cite{ken} which satisfies
\begin{align}
R(B_t(x))-R(x)= Z^x_t +\f{1}{2}\int^t_0 \Delta R(B_s(x))\Id s-L^x_t\>\>\text{ $\mathbb{P}$-a.s.}\label{ecl}
\end{align}
for any $t\geq 0$, $x\in M$, where $Z^x$ is a Brownian motion which starts in $0$, $L^x$ is a continuous nondecreasing process which starts in $0$, and where the integral can be defined since $B(x)$ does not spend time in $\mathrm{Cut}(\mathscr{O})\cup\{\mathscr{O}\}$ (this follows from the well-known fact that $\mathrm{Cut}(\mathscr{O})\cup\{\mathscr{O}\}$ has measure zero; see p.527 in \cite{Ha} for details). For any $x\in M$ let $Y^x:[0,\infty)\times\Omega\to (0,\infty)$ be the uniquely determined maximal solution of 
\begin{align}
\Id Y^x = \Id Z^x+  \f{1}{2}h(Y^x) \Id t,\>\>Y^x_0= \max_{x\in K} R(x)\>(>0), \label{qqpp}
\end{align}
where we remark that the Feller explosion test as formulated in proposition 4.2.2 in \cite{Hsu} can be checked with elementary estimates to prove that $Y^x$ is indeed nonexplosive. Furthermore, (\ref{ecm2}), (\ref{ecl}) and a classical comparison theorem for stochastic differential equations (theorem 1.1 in \cite{ejl}) imply
\begin{align}
 R(B_t(x))\leq Y^x_t\>\>\>\>\text{ $\mathbb{P}$-a.s. for any $t\geq 0$, $x\in K$.}\label{un2} 
\end{align}
Now (\ref{un2}) shows the following uniform estimate in $x$: For any $t\geq 0$ and $r>0$,
\begin{align}
\inf_{x\in K} \mathbb{P} \{t < \zeta(r,x)\}&= \inf_{x\in K} \mathbb{P} \left\lbrace R(B_s(x))< r \text{ for all $s\in [0,t]$}\right\rbrace \nn\\
&\geq \inf_{x\in K} \mathbb{P}\left\lbrace  Y^x_s < r \text{ for all $s\in [0,t]$}\right\rbrace \nn\\
&= \mathbb{P}\left\lbrace  Y^{x^*}_s < r \text{ for all $s\in [0,t]$}\right\rbrace,\label{edh}
\end{align}
where $x^*$ is an arbitrary point in $M$ and where have used uniqueness in law for the pair $(1,h)$. Finally, (\ref{ggd1}) follows from the fact that if $r>\max_{x\in K} R(x)$, then (by the continuity of $Y^{x^*}$) the last term in (\ref{edh}) tends to $1$ as $t\searrow 0$. \vspace{0.5mm}
\hfill$\blacksquare$\vspace{1.1mm}

We will use proposition \ref{kkl} to prove part b) of:

\begin{Proposition}\label{kkl2} {\rm a)}  Let $M$ be stochastically complete and let $|V|\in\mathcal{K}(M)$. Then one has
\begin{align}
\lim_{t\searrow 0}\sup_{x\in M} \mathbb{E}\left[ \left\|\mathbf{1}-\mathscr{V}^{x}_t \right\|^2_x\right]=0.
\end{align}

{\rm b)} Let $M$ be geodesically complete with Ricci curvature bounded from below and let $V$ be such that there is a decomposition $V=V^{(1)}-V^{(2)}$ into potentials $V^{(1)},V^{(2)}\geq 0$ with 
\[
\left|V^{(1)}\right|\in\mathcal{K}_{\mathrm{loc}}(M),\>\>\left|V^{(2)}\right|\in\mathcal{K}(M).
\]  
Then for all compact $K\subset M$ one has
\begin{align}
\lim_{ t\searrow 0}\sup_{x\in K} \mathbb{E}\left[ \left\|\mathbf{1}-\mathscr{V}^{x}_t  \right\|^2_x\right]= 0. \label{qebb}
\end{align}
\end{Proposition}

{\it Proof.} a) By corollary \ref{gs} we have $\mathbb{P}$-a.s. 
\begin{align}
\left\|\mathbf{1}-\mathscr{V}^{x}_t \right\|_x \leq \left(\int^t_0\left\|V(B_s(x))\right\|_{B_s(x)}\Id s\right)^{\f{1}{4}} \mathrm{e}^{\int^t_0\left\|V(B_s(x))\right\|_{B_s(x)}\Id s },
\end{align}
so that Cauchy-Schwarz implies
\begin{align}
\mathbb{E}\left[\left\|\mathbf{1}-\mathscr{V}^{x}_t \right\|^2_x\right] \leq \mathbb{E}\left[\int^t_0\left\|V(B_s(x))\right\|_{B_s(x)}\Id s\right]^{\f{1}{2}} \mathbb{E}\left[ \mathrm{e}^{4\int^t_0\left\|V(B_s(x))\right\|_{B_s(x)}\Id s }\right]^{\f{1}{2}}.\label{cjkj}
\end{align}
Using proposition \ref{xdd}, (\ref{cjkj}) gives 
\begin{align}
\sup_{x\in M} \mathbb{E}\left[ \left\|\mathbf{1}-\mathscr{V}^{x}_t  \right\|^2_x\right]\leq \left(2\mathrm{e}^{C(4|V|)t} \sup_{x\in M} \mathbb{E}\left[\int^t_0\left\|V(B_s(x))\right\|_{B_s(x)}\Id s\right]\right)^{\f{1}{2}}, 
\end{align}
which tends to zero as $t\searrow 0$ by the definition of the Kato class. \vspace{1.2mm}

b) Let $\chi(r,t,x)$ be as in proposition \ref{kkl}. Then we have
\begin{align}
&\sup_{x\in K} \mathbb{E}\left[\Big( 1-  \chi(r,t,x)  + \chi(r,t,x)\Big)  \left\|\mathbf{1}-\mathscr{V}^{x}_t  \right\|^2_x\right]\nn\\
&\leq \sup_{x\in K}  \mathbb{E}\left[1-  \chi(r,t,x) \right]^{\f{1}{2}} \sup_{x\in K}  \mathbb{E}\left[\left\|\mathbf{1}-\mathscr{V}^{x}_t\right\|^2_x \right]^{\f{1}{2}}\nn\\
&\>\>\>\>\>+\sup_{x\in K}\mathbb{E}\left[  \chi(r,t,x)\left\|\mathbf{1}-\mathscr{V}^{x}_t  \right\|^2_x\right]\nn\\
&\leq\left(2+4\mathrm{e}^{C\left(2\left|V^{(2)}\right|\right)t}\right)^{\f{1}{2}} \sup_{x\in K}  \mathbb{E}\left[1-  \chi(r,t,x) \right]^{\f{1}{2}}+\sup_{x\in K}\mathbb{E}\left[  \chi(r,t,x)\left\|\mathbf{1}-\mathscr{V}^{x}_t  \right\|^2_x\right],\label{zbu}
\end{align}
where we have used Cauchy-Schwarz and $1-  \chi(r,t,x)=(1-  \chi(r,t,x))^2$ for the first step and (\ref{gjs}) with $s=0$ for the second step. In view of (\ref{in}), it follows from (\ref{zbu}) that it is sufficient to prove that for (some) $r>\max_{x\in K}\Id (\mathscr{O},x)$ one has
\begin{align}
\lim_{t \searrow 0} \sup_{x\in K}\mathbb{E}\left[  \chi(r,t,x)\left\|\mathbf{1}-\mathscr{V}^{x}_t  \right\|^2_x\right]= 0.\label{edc}
\end{align}

To this end, let $t>0$, $r>\max_{x\in K}\Id (\mathscr{O},x)$ and take a $\Psi\in\mathsf{C}^{\infty}_0(M)$ such that $\Psi=1$ in $\mathrm{K}_r(\mathscr{O})$. We denote with $\mathscr{V}^{\Psi,(x)}$ the pathwise weak solution of (\ref{gg6}) with $V$ replaced with $\Psi V$ and remark that $|\Psi V| \in\mathcal{K}(M)$. Since in $\{  \chi(r,t,x)\ne 0\}$ one $\mathbb{P}$-a.s. has 
\[
\pa^{x,-1}_s V(B_s(x)) \pa^{x}_s= \pa^{x,-1}_s \Psi(B_s(x)) V(B_s(x)) \pa^{x}_s\>\>\text{ for any $0\leq s\leq t$},
\]
expanding $\mathscr{V}^{x}$ and $\mathscr{V}^{\Psi,(x)}$ into path ordered exponentials as in (\ref{pato}) shows
\[
 \mathbb{E}\left[  \chi(r,t,x)\left\|\mathbf{1}-\mathscr{V}^{x}_t  \right\|^2_x\right]=\mathbb{E}\left[ \chi(r,t,x)\left\|\mathbf{1}-\mathscr{V}^{\Psi,(x)}_t  \right\|^2_x\right],
\]
and (\ref{edc}) follows from part a). \vspace{0.5mm}

\hfill$\blacksquare$\vspace{1.1mm}

Now we are prepared to prove theorem \ref{gene}. \vspace{1.1mm}

{\it Proof of theorem \ref{gene}.} The asserted {\it boundedness} follows from setting $s=0$ in (\ref{boundedo}). \vspace{1.2mm}

{\it Continuity:} It follows from remark \ref{uunv} that it is sufficient to prove that for any compact $K\subset M$ and any $f\in \Gamma_{\mathsf{L}^{\infty}}(M,E)\cap\Gamma_{\mathsf{L}^{2}}(M,E)$ one has 
\begin{align}
\lim_{s\searrow 0}\sup_{x\in K} \left\|Q^0_sQ^V_{t-s}f(x)- Q^V_t f(x)\right\|_x=0.\label{d8}
\end{align}
By (\ref{uhgq}),  
\begin{align}
 &\left\|Q^0_sQ^V_{t-s}f(x)- Q^V_t f(x)\right\|_x\nn\\
=&\>\left\|\mathbb{E} \left[ \left(  \mathscr{V}^{x,-1}_s\mathscr{V}^{x}_t-\mathscr{V}^{x}_t\right)   \pa^{x,-1}_t f(B_t(x))\right]\right\|_x\nn\\
=&\>\left\|\mathbb{E} \left[ \left(  \mathbf{1}-\mathscr{V}^{x}_s\right)\mathscr{V}^{x,-1}_s\mathscr{V}^{x}_t   \pa^{x,-1}_t f(B_t(x))\right]\right\|_x\nn\\
\leq &\>\left\|f\right\|_{\infty}\mathbb{E} \left[ \left\|\mathbf{1}-\mathscr{V}^{x}_s\right\|_x\left\|\mathscr{V}^{x,-1}_s\mathscr{V}^{x}_t\right\|_x  \right],
\end{align}
so that using Cauchy-Schwarz with (\ref{gjs}) and proposition \ref{kkl2} b) we get
\begin{align}
& \sup_{x\in K}\left\|Q^0_sQ^V_{t-s}f(x)-Q^V_t f(x)\right\|_x\nn\\
\leq & \> \left\|f\right\|_{\infty}\left(2\mathrm{e}^{ C\left(2\left|V^{(2)}\right|\right)t }\sup_{x\in K} \mathbb{E}\left[\left\| \mathbf{1}-\mathscr{V}^{x}_s\right\|^2_x\right]\right)^{\f{1}{2}}\>\>\text{ $\to 0$, as $s\searrow 0$.}\nn
\end{align}
This completes the proof. \vspace{0.5mm}

\hfill$\blacksquare$\vspace{1.1mm}

We finally remark the following corollary to theorem \ref{gene}, proposition \ref{hyp} and remark \ref{ggf}, which is important for geometric applications:

\begin{Corollary} Let $M$ be geodesically complete with Ricci curvature bounded from below and a positive injectivity radius. Assume furthermore that $V\in\Gamma_{\mathsf{C}^{\infty}}(M,\mathrm{End}(E))$ is such that there is a decomposition $V=V^{(1)}-V^{(2)}$ into potentials $V^{(1)},V^{(2)}\geq 0$ with $\left|V^{(2)}\right|\in\mathcal{K}(M)$. Then for any $t>0$, $f\in\Gamma_{\mathsf{L}^2}(M,E)$, the section 
\begin{align}
 M\longrightarrow E,\>\>x\longmapsto \mathbb{E} \left[ \mathscr{V}^{x}_t  \pa^{x,-1}_t f(B_t(x))\right]\in E_x\label{gg3} 
\end{align}
is $\mathsf{C}^{\infty}$ and bounded. In particular, $\mathrm{e}^{-tH(V)}f$ has a bounded $\mathsf{C}^{\infty}$-representative which is given by (\ref{gg3}).
\end{Corollary}

{\it Proof.} Let $Q^V_tf(x)$ be defined by the right hand side of (\ref{gg3}). By local elliptic regularity, $\mathrm{e}^{-tH(V)}f$ has a $\mathsf{C}^{\infty}$-representative $f^V_t$. The Feynman-Kac formula implies $Q^V_tf(x)=f^V_t(x)$ for a.e. $x$, but since $x\mapsto Q^V_tf(x)$ is continuous (and bounded) by theorem \ref{gene}, we actually have $Q^V_tf(x)=f^V_t(x)$ for {\it all} $x$. \vspace{0.5mm}

\hfill$\blacksquare$ 

\appendix
\numberwithin{theorem}{section}

\section{Some auxiliary Feynman-Kac formulae}\label{scall}

\begin{Proposition}\label{besch} Let $|V|\in\mathsf{L}^{\infty}(M)$. Then for any $f\in \Gamma_{\mathsf{L}^2}(M,E)$, $t\geq 0$, a.e. $x\in M$ one has
\begin{align}
\mathrm{e}^{-t H(V) }f(x)= \mathbb{E} \left[  \mathscr{V}^{x}_t \pa_t^{x,-1} f(B_t(x))1_{\{t<\zeta(x)\}}\right].\label{zxa}
\end{align}
\end{Proposition}

{\it Proof.} If $V$ is smooth and bounded, the formula follows from theorem 9.4 in \cite{Dr}. In the general case, one can use the same arguments as in the proof of theorem 1.1 in \cite{Gue}: Using Friedrichs mollifiers, one can construct a sequence of smooth potentials $(V_n)$ such that (see for example lemma 3.1 in \cite{Gue}; the sequence constructed there is actually smooth)
\begin{align}
|V_n|, |V|\leq C\>\>\text {for all $n$},\>\>|V_n-V|\to 0\>\>\text{ as $n\to\infty$ a.e. in $M$.}\label{sadd} 
\end{align}
Using dominated convergence, this implies 
\[
\left\|V_n f- Vf\right\| \to 0\>\>\text{ as $n\to\infty$ for any $f\in\Gamma_{\mathsf{L}_2}(M,E)$,}
\]
so as $\mathsf{D}(H(0))$ is a common operator core for $H(V)$, $H(V_n)$, we can assume that with an obvious notation one has (see theorem VIII 25, theorem VIII 20 in  \cite{Re})
\[
\mathbb{E} \left[  \mathscr{V}^{x}_{n,t} \pa_t^{x,-1} f(B_t(x))1_{\{t<\zeta(x)\}}\right]\to \mathrm{e}^{-t H(V) }f(x)\>\>\text{ as $n\to\infty$ for a.e. $x\in M$.} 
\]
Now using $|V_n|\leq C$ for all $n$ and expanding  $\mathscr{V}^{x}_{n,t}$ into a path ordered exponential as in (\ref{pato}) implies
\[
\left\|\mathscr{V}^{x}_{n,t}\right\|_x 1_{\{t<\zeta(x)\}}\leq \mathrm{e}^{tC}\>\>\text{ as $n\to\infty$, $\mathbb{P}$-a.s.,}
\]
and combining (\ref{sadd}) with proposition \ref{schlesi} and dominated convergence implies 
\[
\left\|\mathscr{V}^{x}_{n,t}-\mathscr{V}^{x}_{t}\right\|_x 1_{\{t<\zeta(x)\}}\to 0\>\>\text{ as $n\to\infty$, $\mathbb{P}$-a.s.,}
\]
so that (\ref{zxa}) follows from dominated convergence.  \vspace{0.5mm}

\hfill$\blacksquare$\vspace{1.1mm}

Proposition \ref{besch} is a generalization of theorem 1.1 in \cite{Gue} to possibly incomplete $M$'s. When applied to the trivial line bundle, this result directly implies: 

\begin{Corollary}\label{cori2} Let $v\in\mathsf{L}^{\infty}(M)$. Then the following formula holds for any $f\in\mathsf{L}^2(M)$, $t\geq 0$ and a.e. $x\in M$,
\begin{align}
\mathrm{e}^{-t H_0(v)}f(x)=\mathbb{E}\left[\mathrm{e}^{-\int^t_0 v(B_s(x))\Id s}f(B_t(x))1_{\{t<\zeta(x)\}}\right].
\end{align}
\end{Corollary}

Now we can give a proof of theorem \ref{sa2}:\vspace{1.4mm}

{\bf Proof of theorem \ref{sa2}}. The proof follows the strategy of the proof of theorem 6.2 in \cite{Si} (see also \cite{Jo}). We first remark that by writing $f=f_1-f_2+\mathrm{i}f_3-\mathrm{i}f_4$ with $f_j\geq 0$, we can and we will assume $f\geq 0$. We divide the proof into two parts:\vspace{1.2mm}

I){\it (\ref{xx22}) and (\ref{zz4}) hold under the additional assumption $v\geq C$.} \vspace{1.2mm}

Proof: We can assume $C=0$, so $v\geq 0$ and (\ref{xx22}) follows from lemma \ref{Katto0}. Let us define a sequence of potentials $(v_n)\subset \mathsf{L}^{\infty}(M)$ by $v_n:=\min(n,v)$. Then we have  
\begin{align}
&\mathsf{D}(q_{H_0(v)})=\mathsf{D}(q_{-\Delta/2})\cap \left.\Big\{\psi\right|v^{\f{1}{2}}\psi\in \mathsf{L}^2(M)\Big\},\label{f1}\\
&q_{H_0(v)}(\psi)=q_{-\Delta/2}(\psi)+\int_M v(x) |\psi(x)|^2 \mathrm{vol}(\Id x),\label{f2}
\end{align}
and for any $n$ it holds that $\mathsf{D}(q_{H_0(v_n)})=\mathsf{D}(q_{-\Delta/2})$ with
\begin{align}
q_{H_0(v_n)}(\psi)=q_{-\Delta/2}(\psi)+\int_M v_n(x) |\psi(x)|^2 \mathrm{vol}(\Id x),\label{f3}
\end{align}
and it follows from $0\leq v_n\leq v_{n+1}\leq v$, $v_n\to v$ a.e. in $M$ as $n\to\infty$ and monotone convergence of integrals and convergence of monotonely increasing quadratic forms (see for example theorem 12.2.2 in \cite{Jo}), that we may assume   
\begin{align}
\lim_{n\to\infty}\mathrm{e}^{-t H_0(v_n)}f(x)= \mathrm{e}^{-t H_0(v)}f(x)\>\> \text{  for a.e. $x\in M$.} \label{f4}
\end{align}
Corollary \ref{cori2} implies
\begin{align}
\mathrm{e}^{-t H_0(v_n)}f(x)= \mathbb{E}\left[\mathrm{e}^{-\int^t_0 v_n(B_s(x))\Id s }f(B_t(x))1_{\{t<\zeta(x)\}}\right]\>\> \text{  for a.e. $x$,}\label{e33s}
\end{align}
so that  
\begin{align}
\mathrm{e}^{-t H_0(v)}f(x)=\lim_{n\to\infty}\mathbb{E}\left[\mathrm{e}^{-\int^t_0 v_n(B_s(x))\Id s }f(B_t(x))1_{\{t<\zeta(x)\}}\right]\>\> \text{  for a.e. $x$.} \label{ee44}
\end{align}
Next, one gets from combining $\mathrm{e}^{-\int^t_0 v^{(1)}(B_s(x))\Id s}f(B_t(x))1_{\{t<\zeta(x)\}}\in \mathsf{L}^1(\mathbb{P})$ with
\begin{align}
\mathrm{e}^{-\int^t_0 v^{(1)}(B_s(x))\Id s}f(B_t(x)) \geq \mathrm{e}^{-\int^t_0 v_n(B_s(x))\Id s}f(B_t(x))\geq \mathrm{e}^{-\int^t_0 v(B_s(x))\Id s}f(B_t(x)) \nn
\end{align}
$\mathbb{P}$-a.s. in $\{t<\zeta(x)\}$ and a generalized abstract convergence theorem for integrals (theorem 12.2.6 in \cite{Jo}), that the right-hand side of (\ref{ee44}) is equal to
\[
\mathbb{E}\left[\exp\left(-\lim_{n\to\infty}\int^t_0 v_n(B_s(x))\Id s \right)f(B_t(x))1_{\{t<\zeta(x)\}}\right]  \>\> \text{  for all $x$,}
\]
and $\int^t_0 v_n(B_s(x))\Id s \to \int^t_0 v(B_s(x))\Id s$, $\mathbb{P}$-a.s. in $\{t<\zeta(x)\}$ as $n\to\infty$, follows from monotone convergence. \\

II) {\it (\ref{xx22}) and (\ref{zz4}) hold in the general case.} \vspace{1.2mm}

Proof: Again, (\ref{xx22}) follows from lemma \ref{Katto0}. It remains to prove (\ref{zz4}): We define $v_n:=\max(-n,v)$, so that each $v_n$ is a bounded from below, locally integrable potential and $ v_n\geq v_{n+1}\geq v$, $v_n\to v$ a.e. in $M$ as $n\to\infty$. One has $\mathsf{D}(q_{H_0(v_n)})=\mathsf{D}(q_{H_0(v)})$ and
\begin{align}
&q_{H_0(v)}(\psi)=q_{-\Delta/2}(\psi)+\int_M v(x) |\psi(x)|^2 \mathrm{vol}(\Id x),\\
&q_{H_0(v_n)}(\psi)=q_{-\Delta/2}(\psi)+\int_M v_n(x) |\psi(x)|^2 \mathrm{vol}(\Id x).
\end{align}
Furthermore, one can use the above cited generalized convergence theorem for integrals and convergence of monotonely decreasing quadratic forms (the latter by subtracting $\mathscr{E}_{H_0(v)}$ if necessary) to see that we may assume (\ref{f4}) again. By I), we also have (\ref{e33s}) now. It remains to prove 
\begin{align}
\mathbb{E}\left[\mathrm{e}^{-\int^t_0 v_n(B_s(x))\Id s }f(B_t(x))1_{\{t<\zeta(x)\}}\right]\to \mathbb{E}\left[\mathrm{e}^{-\int^t_0 v(B_s(x))\Id s }f(B_t(x))1_{\{t<\zeta(x)\}}\right]\label{ff}
\end{align}
as $n\to\infty$ for a.e. $x$. Noting that by I) for a.e. $x$ one has
\[
\mathrm{e}^{-\int^t_0 v^{(1)}(B_s(x))\Id s}f(B_t(x))1_{\{t<\zeta(x)\}}\in \mathsf{L}^1(\mathbb{P}),\>\> \int^t_0|v^{(1)}(B_{s}(x))|\Id s<\infty,
\]
the latter $\mathbb{P}$-a.s. in $\{t<\zeta(x)\}$, we can use theorem 12.2.6 in \cite{Jo} twice to see that (\ref{ff}) holds, which completes the proof. \vspace{0.5mm}\hfill$\blacksquare$

\section{Proof of proposition \ref{as11}}\label{th}

Let us first note the following simple fact: For any $r\geq 1$ and $x\in M$ one has
\begin{align}
 \left\|p_t(x,\bullet)\right\|_r\leq C(t,r):=C_t^{1-\f{1}{r}}.\label{hilfe}
\end{align}
Throughout, let $h\in \mathsf{L}^p(M)$.\vspace{1.2mm}
 
Case $1<p<q<\infty$: Let $r$ be given as $1-1/r=1/p-1/q$. Applying Hölder's inequality with the exponents 
\[
q_1=q,\>\>q_2=\f{r}{1-\f{r}{q}},\>\>q_3=\f{p}{1-\f{p}{q}} 
\]
gives that $\left\|P_th\right\|^q_q$ is
\begin{align}
&\leq \int_M\left( \int_M \left(  p_t(x,y)^r |h(y)|^p\right)^{\f{1}{q}}p_t(x,y)^{1-\f{r}{q}}|h(y)|^{1-\f{p}{q}}    \mathrm{vol}(\Id y)\right) ^q \mathrm{vol}(\Id x)\nn\\
&\leq \int_M \left(\int_M p_t(x,y)^r|h(y)|^p \mathrm{vol}(\Id y) \right) \left(\int_M p_t(x,y)^r\mathrm{vol}(\Id y)  \right)^{\f{q}{r}\left(1-\f{r}{q} \right)}\nn\\
&\>\>\>\>\times \left( \int_M |h(y)|^p\mathrm{vol}(\Id y) \right)^{\f{q}{p}\left(1-\f{p}{q} \right) } \mathrm{vol}(\Id x),\nn
\end{align}
so that using Fubini's theorem and (\ref{hilfe}),
\begin{align}
&\left\|P_th\right\|^q_q\leq   C(t,r)^{ q\left(1-\f{r}{q} \right)}\left\|h\right\|^{q\left(1-\f{p}{q} \right) }_p \int_M |h(y)|^p\int_M p_t(x,y)^r \mathrm{vol}(\Id x) \mathrm{vol}(\Id y)\nn\\
&\leq  C(t,r)^{q}\left\|h\right\|^{q }_p=  C_t^{q\left( \f{1}{p}-\f{1}{q}\right) }\left\|h\right\|^{q }_p.
\end{align}

Case $1<p=q<\infty$: One has 
\begin{align}
&\left\|P_th\right\|^p_p \leq \int_M \left(\int_M p_t(x,y) |h(y)|\mathrm{vol}(\Id y) \right)^p\mathrm{vol}(\Id x)\nn\\
&\leq \int_M \int_M  |h(y)|^p p_t(x,y)\mathrm{vol}(\Id y)\mathrm{vol}(\Id x)\label{trick}\\
&= \int_M \int_M   p_t(x,y)\mathrm{vol}(\Id x)|h(y)|^p\mathrm{vol}(\Id y)\nn\\
&\leq \left\|h\right\|^{p }_p,
\end{align}
where we have applied the Hölder inequality to the finite measure $\mu(\Id y)=p_t(x,y) \mathrm{vol}(\Id y)$ for the second inequality.\vspace{1.2mm}

Case $1<p<q=\infty$: This works with the same argument that has been used for the inequality (\ref{trick}). \vspace{1.2mm}

Case $1=p<q<\infty$: One has 
\begin{align}
 \left\| P_th\right\|^q_q \leq \int_M\left( \int_M \left(  p_t(x,y)^q |h(y)|\right)^{\f{1}{q}}|h(y)|^{1-\f{1}{q}}    \mathrm{vol}(\Id y)\right) ^q \mathrm{vol}(\Id x).
\end{align}
Applying the Hölder inequality with the exponents 
\[
q_1=q,\>\> q_2=\f{1}{1-\f{1}{q}}
\]
gives
\begin{align}
 \left\| P_th\right\|^q_q \leq \left\|h\right\|^{q-1 }_1\int_M\int_M  p_t(x,y)^q |h(y)|\mathrm{vol}(\Id y) \mathrm{vol}(\Id x),
\end{align}
so that the Fubini theorem and (\ref{hilfe}) imply
\[
 \left\|P_t h\right\|^q_q\leq C_t^{q\left( 1-\f{1}{q}\right) }\left\|h\right\|^q_1.
\]

The cases $p=q=\infty$ and $p=q=1$ and $p=1$,$q=\infty$ are trivial. \vspace{0.5mm}

\hfill$\blacksquare$

\section{Some inequalities}

Let $\IHH$ be a finite dimensional complex or real Hilbert space with scalar product $\left\langle \bullet,\bullet\right\rangle$ and the corresponding norm $\left\|\bullet\right\|$. The induced operator norm will be denoted with the same symbol. If $0\leq a<b\leq\infty$, $F\in \mathsf{L}^1_{\mathrm{loc}}([a,b),\ILL(\IHH))$, then a standard use of the Banach fixed point theorem shows that there is a unique weak (= locally absolutely continuous) solution $Y:[a,b)\to  \ILL(\IHH)$ of the ordinary initial value problem
\begin{align}
\f{\Id}{\Id s} Y(s)=Y(s)F(s),\>\>Y(a)=\mathbf{1}.\label{cck}
\end{align}
It is easily seen that $Y$ is invertible with 
\[
\f{\Id}{\Id s} Y^{-1}(s)=-F(s)Y^{-1}(s),\>\>Y(a)=\mathbf{1}.
\]

\begin{Proposition}\label{absc} Let $F$ and $Y$ be as above.\vspace{1.2mm}

{\rm a)} For any $a\leq t< b$,
\[
\left\|Y(t)\right\|\leq \mathrm{e}^{\int^t_a \left\|F(s)\right\|\Id s}.
\]

{\rm b)} Let $a=0$. For any $0\leq t<b$,
\begin{align}
\left\|Y(t)-\mathbf{1}\right\|\leq \mathrm{e}^{\int^t_0 \left\|F(s)\right\|\Id s}. \label{s2y}
\end{align}

{\rm c)} Let $a\leq t<b$, assume that $F(s)$ is Hermitian for a.e. $s\in [a,t]$ and that there exists a real-valued function $c\in\mathsf{L}^1[a,t]$ such that for all $v\in\IHH$ one has
\[
\left\langle F(s) v, v\right\rangle\leq c(s) \left\|v\right\|^2\>\>\text{ for a.e. $s\in[a,t]$.}
\]
Then one has
\[
\left\|Y(t)\right\|\leq  \mathrm{e}^{\int^t_a c(s) \Id s}.
\]

{\rm d)} Let $a=0$, $0\leq t_1\leq t_2<b$ and assume that $F(s)$ is Hermitian for a.e. $s\in [0,t_2]$ and that there exists a real-valued function $c\in\mathsf{L}^1[0,t_2]$ such that for all $v\in\IHH$ one has
\[
\left\langle F(s) v, v\right\rangle\leq c(s) \left\|v\right\|^2\>\>\text{ for a.e. $s\in[0,t_2]$.}
\]
Then one has
\[
\left\|Y^{-1}(t_1)Y(t_2)\right\|\leq  \mathrm{e}^{\int^{t_2}_{t_1} c(s) \Id s}.
\]
\end{Proposition}

{\it Proof.} a) This is an obvious analogue of B.1.(a) in \cite{Gue}.\vspace{1.2mm}

b) This follows easily from expanding $Y$ into the path ordered exponential
\begin{align}
Y(t)=\mathbf{1}+\int_{0\leq s_1\leq \dots\leq s_k\leq t} F(s_1)\dots F(s_k) \Id s_1\dots \Id s_k.\label{pato}
\end{align}

c) This is an analogue of proposition B.1.(b) in \cite{Gue}. \vspace{1.2mm}

d) This follows from part c), by noting that for fixed $t_1$, the function 
\[
Y^{-1}(t_1)Y(\bullet):[t_1,b)\longrightarrow \ILL(\IHH)
\]
is the solution of (\ref{cck}) with $a=t_1$.\vspace{0.5mm}

\hfill$\blacksquare$\vspace{1.1mm}

\begin{Proposition}\label{schlesi} Let $F_1$, $F_2\in \mathsf{L}^1_{\mathrm{loc}}([a,b),\ILL(\IHH))$ and let 
\[
Y_1, Y_2:[a,b)\longrightarrow  \ILL(\IHH)
\]
be the unique solutions of the ordinary initial value problems
\[
\f{\Id}{\Id s} Y_j(s)= Y_j(s)F_j(s),\>\>Y_j(a)=\mathbf{1}\>\>\text{ for }j=1,2.
\]
The following inequality holds for all $a\leq t<b$,
\begin{align}
\left\|Y_1(t)-Y_2(t)\right\|\leq   \> & \mathrm{e}^{2\int^t_a \left\|F_{1}(s )\right\|\Id s+\int^t_a \left\|F_{2}(s)\right\|\Id s }\int^t_a  \left\|F_{1}(s)-   F_{2}(s)\right\|\Id s.\label{s2x}
\end{align}
\end{Proposition}

{\it Proof.} This is proposition B.2 in \cite{Gue}.\vspace{0.5mm}

\hfill$\blacksquare$\vspace{1.1mm}

\begin{Corollary}\label{gs} Let $F$ and  $Y$ be as above and let $a=0$. Then for any $ 0\leq t<b$ and any $p\geq 1$ one has
\begin{align}
\left\|Y(t)-\mathbf{1}\right\|\leq \left(\int^t_0\left\|F(s)\right\|\Id s\right)^{\f{1}{p}} \mathrm{e}^{\int^t_0 \left\|F(s)\right\|\Id s }.\label{ccd}
\end{align}
\end{Corollary}

{\it Proof.} If $\int^t_0\left\|F(s)\right\|\Id s\leq 1$, then (\ref{ccd}) follows from applying (\ref{s2x}) with $F_1=0$. If $\int^t_0\left\|F(s)\right\|\Id s> 1$, then (\ref{ccd}) follows from (\ref{s2y}). \vspace{0.5mm}

\hfill$\blacksquare$ 
 
 \section{Some Hilbert space facts}

We collect some well-known Hilbert space facts in the following theorem. 

\begin{Theorem} Let $\IHH=(\IHH,\left\langle \bullet,\bullet\right\rangle)$ be a Hilbert space and let $\left\|\bullet\right\|$ be the corresponding norm. \vspace{1mm}

{\rm a)} Let $B\in\ILL(\IHH)$ be self-adjoint. Then one has  
\begin{align}
\max\sigma\left( B\right) = \sup \left.\Big\{ \left\langle Bf,f\right\rangle\right|f\in\IHH,\left\|f\right\|=1\Big\}.\label{ap0}
\end{align}

{\rm b)} Let $H\geq c_1$ be self-adjoint and let $q_H$ be the quadratic form corresponding to $H$. Then for any number $c_2 \leq  c_1$ one has 
\begin{align} 
\mathsf{D}(q_H)=\mathsf{D}\left( (H-c_2)^{\f{1}{2}}\right) ,\>\>q_H(f)=\left\|(H-c_2)^{\f{1}{2}}f\right\|^2+c_2\left\|f\right\|^2. \label{ap1}
\end{align}
Furthermore, with $\mathscr{E}_H:=\min \sigma(H)$ it holds that 
\begin{align}
\mathscr{E}_H=\inf \left.\Big\{ q_H(f)\right|f\in\mathsf{D}(q_H),\left\|f\right\|=1\Big\}\label{ap2}
\end{align}
and 
\begin{align}
\max \sigma(\mathrm{e}^{-H})=\mathrm{e}^{- \mathscr{E}_H}. \label{ap3}
\end{align}
Finally, if $c_1=0$, that is $H\geq 0$, then
\begin{align}
 &\mathsf{D}(q_H)=\left\lbrace f\left|f\in \IHH,\> \lim_{t\searrow 0}\left\langle \f{f-\mathrm{e}^{-t H}f}{t},f\right\rangle<\infty \right\rbrace \right., \nn\\
&q_H(f)= \lim_{t\searrow 0}\left\langle \f{f-\mathrm{e}^{-t H}f}{t},f\right\rangle. \label{ap4}
\end{align}
\end{Theorem}

{\it Proof.} a) This follows from theorem 2.19 in \cite{teschl}. \\
b) (\ref{ap1}) can be found on p. 332 in \cite{katze2}, (\ref{ap2}) is included in Satz 8.27 in \cite{weidmann}, (\ref{ap3}) can be found on p. 322 in \cite{weidmann}, and (\ref{ap4}) follows from applying (\ref{ap1}) with $c_j=0$ and the spectral calculus. \hfill$\blacksquare$\vspace{2mm}

{\bf Acknowledgements.} The author would like to thank A. Thalmaier, K. Kuwada and R. Philipowski for helpful discussions on proposition \ref{kkl4} and proposition \ref{kkl}. The reasearch has been financially supported by the Bonner Internationale Graduiertenschule and the SFB 647: Raum - Zeit - Materie.


\begin{thebibliography}{99}



\bibitem{Br} Braverman, M. \&  Milatovich, O. \& Shubin, M.: {\it Essential self-adjointness of Schrödinger-type operators on manifolds.} Russian Math. Surveys  57  (2002),  no. 4, 641--692.


\bibitem{Bro1} Broderix, K. \& Hundertmark, D. \& Leschke, H.: {\it Continuity properties of Schrödinger semigroups with magnetic fields.} Reviews in Mathematical Physics 12 (2000), 181--225.


\bibitem{Bru} Brüning, J. \& Geyler, V. \& Pankrashkin, K.: {\it Continuity properties of integral kernels associated with Schrödinger operators on manifolds.}  Ann. Henri Poincaré  8  (2007),  no. 4, 781--816.

\bibitem{chung} Chung, K.L. \& Xin, Z.Z.: {\it From Brownian motion to Schrödinger's equation.} Grundlehren der Mathematischen Wissenschaften, 312. Springer-Verlag, Berlin, 1995.



\bibitem{dollard} Dollard, J.D. \& Friedman, C.N.: {\it Product Integration.} Addison-Wesley, 1979.


\bibitem{Dr} Driver, B.K. \& Thalmaier, A.: {\it Heat equation derivative formulas for vector bundles.}  J. Funct. Anal.  183  (2001),  no. 1, 42--108.

\bibitem{El} Elworthy, K.D.: {\it  Stochastic Differential Equations on Manifolds} London Mathematical Society Lecture Note Series 70, Cambridge University Press, Cambridge, 1982.


\bibitem{enciso} Enciso, A.: {\it Coulomb Systems on Riemannian Manifolds and Stability of Matter.} Ann. Henri Poincare 12 (2011), 723--741.



\bibitem{Gue} Güneysu, B.: {\it The Feynman-Kac formula for Schrödinger operators on vector bundles over complete manifolds.} J. Geom. Phys. 60 (2010) 1997--2010. 

\bibitem{batu} Güneysu, B.: {\it Kato's inequality and form boundedness of Kato potentials on arbitrary Riemannian manifolds.} Preprint. 

\bibitem{babba} Güneysu, B.: {\it On the Feynman Kac formula for Schrödinger semigroups on vector bundles.} PhD thesis, Bonn (2011).

\bibitem{Gue2} Güneysu, B.: {\it Multiplicative matrix-valued functionals and continuity properties of semigroups corresponding to differential operators with matrix coefficients.} J. Math. Anal. Appl. 380 (2011), 709--725.

\bibitem{G6} Güneysu, B.: {\it Nonrelativistic Hydrogen type stability problems on nonparabolic 3-manifolds}. To appear in Annales Henri Poincaré.

 
\bibitem{Ha} Hackenbroch, W. \& Thalmaier, A.: {\it Stochastische Analysis.} B. G. Teubner, 1994.


\bibitem{hemp}  Hempel, R. \& Voigt, J.: {\it The spectrum of a Schrödinger operator in $L_p(\IR^{\nu})$ is $p$-independent.} Comm. Math. Phys.  104  (1986),  no. 2, 243--250.


\bibitem{Hess1} Hess, H. \& Schrader, R. \& Uhlenbrock, D.A.: {\it Domination of semigroups and generalization of Kato's inequality.} Duke Math. J. 44, Number 4 (1977), 893--904.

\bibitem{Hsu} Hsu, E.: {\it Stochastic Analysis on Manifolds.} AMS, 2002.

\bibitem{Hsu2} Hsu, E.: {\it Heat semigroup on a complete Riemannian manifold.} Annals of Probability 17, Number 3 (1989), 1248--1254. 

\bibitem{ejl} Ikeda, N. \& Watanabe, S.: {\it A comparison theorem for solutions of stochastic differential equations and its applications.} Osaka J. Math. 14 (1977), no. 3, 619--633.


\bibitem{Jo} Johnson, G.W. \& Lapidus, M. L.: {\it The Feynman integral and Feynman's operational calculus.} The Clarendon Press, Oxford University Press, 2000.


\bibitem{katze2} Kato, T.: {\it Perturbation theory for linear operators.} Die Grundlehren der mathematischen Wissenschaften, Band 132, Springer-Verlag New York, Inc., New York 1966.

 
\bibitem{ken} Kendall, W.: {\it The radial part of Brownian motion on a manifold: a semimartingale property.}  Ann. Probab.  15  (1987),  no. 4, 1491--1500.


\bibitem{ku2} Kuwae, K. \& Takahashi, M.: {\it Kato class functions of Markov processes under ultracontractivity.} Potential theory in Matsue, 193--202, Adv. Stud. Pure Math., 44, Math. Soc. Japan, Tokyo, 2006.


\bibitem{kt} Kuwae, K. \& Takahashi, M.: {\it Kato class measures of symmetric Markov processes under heat kernel estimates.}  J. Funct. Anal. 250 (2007), no. 1, 86--113.

\bibitem{lenz} Lenz, D. \& Keller, M. \& Vogt, H. \& Wojciechowski, R.: {\it Note on basic features of large time behaviour of heat kernels.} Preprint.

\bibitem{meyer} Meyer, P.-A.: {\it Martingales and stochastic Integrals I.} Lecture Notes in Mathematics. Vol. 284. Springer-Verlag, Berlin-New York, 1972.  

\bibitem{qian} Qian, Z.: {\it On conservation of probability and the strong Feller property.} Ann. Probab. 24, no. 1 (1996), 280--292.


\bibitem{Re} Reed, M. \& Simon, B.: {\it Methods of modern mathematical physics. I. Functional analysis.} Academic Press, Inc., 1972.


\bibitem{Re4} Reed, M. \& Simon, B.: {\it Methods of modern mathematical physics. IV. Analysis of operators.} Academic Press, Inc., 1978.


\bibitem{Rev} Revuz, D. \&  Yor, M.: {\it Continuous martingales and Brownian motion.} Grundlehren der Mathematischen Wissenschaften, 293. Springer-Verlag, Berlin, 1991. 



\bibitem{Si0} Simon, B.: {\it Schrödinger semigroups.} Bull. Amer. Math. Soc. (N.S.) Volume 7, Number 3 (1982), 447--526. 

\bibitem{Si} Simon, B.: {\it Functional integration and quantum physics.} Academic Press. Inc., 1979.

\bibitem{Sz} Sznitman, A.S.: {\it Brownian motion, obstacles and random media.} Springer, Berlin, 1998.

\bibitem{teschl} Teschl, G.: {\it Mathematical methods in quantum mechanics. With applications to Schrödinger operators.} Graduate Studies in Mathematics, 99. American Mathematical Society, Providence, RI, 2009.


\bibitem{weidmann} Weidmann, J.: {\it Lineare Operatoren in Hilberträumen. Teil 1.} B. G. Teubner, Stuttgart, 2000.

\end{thebibliography}
\end{document}